\documentclass[12pt, oneside, a4paper]{article}
\usepackage{enumitem}
\usepackage{amsmath}
\usepackage{amssymb}
\usepackage{amsthm}
\usepackage{bbm}
\usepackage{amsfonts}
\usepackage{mathtools}
\usepackage{tikz-cd}
\usetikzlibrary{shapes,arrows}
\mathtoolsset{showonlyrefs}

\DeclareMathOperator*{\argmax}{arg\,max}
\DeclareMathOperator*{\argmin}{arg\,min}

\usepackage{etoolbox}

\usepackage[nottoc,notlot,notlof]{tocbibind}

\usepackage{listings}
\usepackage{color}
\definecolor{mygreen}{RGB}{28,172,0}
\definecolor{mylilas}{RGB}{170,55,241}

\newtheorem*{theorem*}{Theorem}
\newtheorem*{remark*}{Remark}

\usepackage{amsthm}

\theoremstyle{remark}
\newtheorem{remark}{Remark}[subsection]

\usepackage{xpatch}
\makeatletter
\xpatchcmd{\@thm}{\thm@headpunct{.}}{\thm@headpunct{}}{}{}
\makeatother

\usepackage{hyperref}

\usepackage[british]{babel}

\usepackage{caption}
\usepackage{subcaption}
\usepackage{graphicx}
\usepackage[justification=centering]{caption}
\usepackage{float}
\usepackage{longtable}
\usepackage{tabularx}
\newcolumntype{Y}{>{\centering\arraybackslash}X}
\usepackage{mathtools}
\usepackage{enumitem}

\usepackage{titlesec}
\titleformat{\chapter}[display]
{\normalfont\huge\bfseries}{}{0pt}{\Huge}
\titlespacing*{\chapter} {0pt}{20pt}{40pt}

\usepackage[thinlines]{easytable}
\usepackage{array}

\usepackage[toc]{appendix}

\usepackage{authblk}

\title{Tomography for Plasma Imaging: a Unifying Framework for Bayesian Inference}
\author[1,2,*]{D. Hamm\thanks{*Corresponding author: daniele.hamm@epfl.ch}}
\author[1]{C. Theiler}
\author[2]{M. Simeoni}
\author[1]{B. P. Duval}
\author[2]{\\T. Debarre}
\author[1]{L. Simons}
\author[2]{J. R. Queralt}
\affil[1]{EPFL-SPC, CH-1015 Lausanne, Switzerland}
\affil[2]{EPFL-Center for Imaging, CH-1015 Lausanne, Switzerland}
\date{}                    
\begin{document}

\makeatletter
  \let\orig@fnsymbol\@fnsymbol
  \def\@fnsymbol#1{}
  \maketitle
  \addtocounter{footnote}{-1}
  \let\@fnsymbol\orig@fnsymbol
\makeatother

\vspace{-1cm}
\begin{abstract}
Plasma diagnostics often employ computerized tomography to estimate emissivity profiles from a finite, and often limited, number of line-integrated measurements. Decades of algorithmic refinement have brought considerable improvements, and led to a variety of employed solutions. These often feature an underlying, common structure that is rarely acknowledged or investigated. In this paper, we present a unifying perspective on sparse-view tomographic reconstructions for plasma imaging, highlighting how many inversion approaches reported in the literature can be naturally understood within a Bayesian framework. In this setting, statistical modelling of acquired data leads to a likelihood term, while the assumed properties of the profile to be reconstructed are encoded within a prior term. Together, these terms yield the posterior distribution, which models all the available information on the profile to be reconstructed. We show how credible reconstructions, uncertainty quantification and further statistical quantities of interest can be efficiently obtained from noisy tomographic data by means of a stochastic gradient flow algorithm targeting the posterior. This is demonstrated by application to soft x-ray imaging at the TCV tokamak. We validate the proposed imaging pipeline on a large dataset of generated model phantoms, showing how posterior-based inference can be leveraged to perform principled statistical analysis of quantities of interest. Finally, we address some of the inherent, and thus remaining, limitations of sparse-view tomography. All the computational routines used in this work are made available as open access code.
\end{abstract}

\newpage

\section{Introduction}\label{introduction}
Fusion devices like tokamaks and stellarators are equipped with a large set of diagnostics \cite{tfr_tokamak_1978}, which are crucial for operation and research purposes. Some of these diagnostics measure the electromagnetic radiation over different photon energies from the devices. Since fusion plasmas can typically be considered optically thin at these photon energies, detectors located in diagnostic ports around the vessel measure the plasma emission integrated over the volume subtended to the detectors \cite{ingesson_chapter_2008}. For sufficiently narrow viewing volumes, detectors provide, to a first approximation \cite{ingesson_chapter_2008, vezinet_mesure_2013, jardin_use_2017}, line-integrated measurements along their so-called lines of sight.
These measurements, also called projections, can then be used to estimate the spatial distribution of the emissivity.
This is achieved using computerised tomography (CT) techniques \cite{ingesson_chapter_2008, kak_principles_2001}. CT is ubiquitous in industrial and medical applications. These typically involve a large number of views (often $10^5-10^6$) \cite{kak_principles_2001}, with lines of-sight evenly distributed around the target being reconstructed to optimize coverage. Diagnostics probing fusion plasmas, in contrast, typically provide $\approx 10^2$ measurements acquired by detectors unevenly distributed around the vessel, due to technical constraints and design choices. Few viewing angles are available, since detectors are usually grouped in camera modules positioned in diagnostic ports. Therefore, tomographic reconstruction from fusion devices' data is often a sparse-view, limited-angle tomography problem. This data scarcity negatively affects the overall quality of plasma emissivity reconstruction, limiting the achievable spatial resolution and increasing artifact proneness \cite{ingesson_chapter_2008}.\\
A wide range of reconstruction techniques has been proposed and tested by the plasma imaging community. In general, most approaches attempt to obtain reconstructions that explain well the tomographic measurements while also promoting some form of smoothness in the reconstruction. Promoting smoothness helps mitigate the scarcity of available data, by driving the reconstruction towards more physically realistic solutions. We refer to \cite{ingesson_chapter_2008, mlynar_current_2019} for reviews of the most relevant efforts published to date. Smoothness is typically enforced using some form of regularization, expressed from a range of principles. Initial constrained/unconstrained optimization approaches were based upon Maximum Entropy \cite{ertl_maximum_1996}, Minimum Fisher Information (MFI) \cite{anton_x-ray_1996} and generalised Tikhonov regularization penalizing sharp derivatives \cite{anton_x-ray_1996}; these approaches have been further refined to leverage knowledge of the magnetic field equilibrium \cite{fuchs_2d_1994, ingesson_soft_1998, odstrcil_modern_2012}, which often improves reconstruction quality. In the case of the Maximum Likelihood algorithm \cite{craciunescu_maximum_2008}, instead, smoothness is promoted through filtering techniques; a form of uncertainty quantification (UQ) was later added \cite{craciunescu_evaluation_2016, craciunescu_maximum_2018}. Gaussian Process Tomography approaches, fully Bayesian, have also been proposed and successfully applied \cite{svensson_nonparametric_2011, li_bayesian_2013, wang_incorporating_2018}; their attractiveness is that the resulting inversions and UQ are available in closed-form. All these techniques rely upon common fundamental Bayesian principles, as will be discussed herein.\\
Bayesian inference is becoming increasingly popular in nuclear fusion research \cite{pavone_machine_2023}. It enables integrated data analysis approaches that combine measurements from multiple diagnostics \cite{fischer_bayesian_2003, svensson_large_2007}, and has been successfully applied to the estimation of current distributions \cite{svensson_current_2008, kwak_bayesian_2022}, electron density and temperature profiles \cite{fischer_integrated_2010}, and divertor characteristics \cite{bowman_development_2020}, to name but a few.\\
In this paper, we focus on the application of Bayesian inference to emission tomography, with soft x-ray (SXR) imaging \cite{sheikh_radcam-radiation_2022} on the TCV tokamak \cite{duval_experimental_2024} as a specific application. However, the proposed framework is completely general and the discussion readily extends to any diagnostic employing tomography (bolometry \cite{sheikh_radcam-radiation_2022}, visible \cite{perek_measurement_2021}, hard x-ray \cite{gnesin_suprathermal_2008}, gamma-ray \cite{kiptily_gamma-ray_2002}, $\ldots$), on TCV and other machines.\\
\vspace{-0.2cm}\\
Our main contributions in this paper are as follows.
\begin{itemize}[leftmargin=1em]
    \item We provide a unifying picture of most state of the art plasma emission tomography techniques, highlighting how Bayesian inference principles lie behind most works to date.
    \vspace{-0.2cm}
    \item We propose a general Bayesian framework that promotes a novel perspective on plasma tomography, offering a highly adaptable and statistically principled environment which, by design, encompasses previous efforts and can easily be built upon. This framework paves the way for the application to plasma imaging of the many recent advances made by the broader computational imaging community, which include, for example, priors promoting sparsity rather than smoothness and powerful data-driven priors.
    \vspace{-0.7cm}
    \item We describe how to obtain, within the proposed framework, estimates of quantities of interest equipped with uncertainty quantification metrics, by means of well-understood, robust, and efficient inference algorithms. In particular, we propose for plasma tomography the unadjusted Langevin algorithm, a Bayesian inference algorithm that is gaining traction in the imaging community.
   \vspace{-0.2cm}
    \item We illustrate a concrete application of the entire workflow to SXR imaging on TCV, analyzing the results of a phantom-based study.
     \vspace{-0.2cm}
   \item In an effort towards reproducibility and reusability, we provide full open access to the computational routines and data used in this study, making them available to the community as a support for further tomography development and/or validation studies.
\end{itemize}
The article is organized as follows. In Section \ref{preliminary}, we introduce some preliminary notions on inference and optimization, describing the Bayesian framework and its connection to classical regularization-based approaches.
In Section \ref{posterior}, we address the problem of extracting information from the posterior probability distribution arising from a Bayesian approach. To this end, we introduce the stochastic gradient flow algorithm \cite{roberts_exponential_1996, durmus_efficient_2018}, which is central to this work. In Section \ref{unifying}, we formalize the plasma tomography inverse problem and review popular approaches to plasma imaging, showing how the Bayesian framework naturally provides a unifying view on these techniques.
In Section \ref{application}, we apply the proposed framework to SXR imaging on TCV. We first describe the generation of a large dataset of model phantoms. The pipeline is then tested on these phantoms. We report the obtained results, demonstrating how the Bayesian framework enables advanced statistical analysis of the tomographic reconstruction and of some physical quantities derived from it. We also comment on some of the inherent limitations posed by sparse-view tomography. Finally, in Section \ref{conclusion}, we summarize and discuss the obtained results, and conclude describing how we plan to build upon this work in future research.

\section{Inverse Problems: Bayesian Perspective}\label{preliminary}
Let us consider the linear inverse problem of recovering the signal $\mathbf{x}\in\mathbb{R}^N$ from the noisy measurements $\mathbf{y}\in\mathbb{R}^M$, with
\begin{equation}\label{ip_data}
    \mathbf{y} = \mathbf{T}\mathbf{x} + \boldsymbol{\varepsilon},
\end{equation}
where $\mathbf{T}\in\mathbb{R}^{M\times N}$ is a linear forward model describing the acquisition system and $\boldsymbol{\varepsilon}\in\mathbb{R}^M$ is the noise corrupting the measurements. Unfortunately, most inverse problems encountered in science are ill-posed \cite{bertero_introduction_2020}, i.e., there may be no solution, or a solution may exist but not be unique, or small perturbations of the data $\mathbf{y}$ may greatly affect the solution. In particular, when the problem is under-determined ($N\!>\!M$) and $\mathbf{T}$ is full rank, there are an infinite number of solutions $\mathbf{x}^*$ that perfectly satisfy the data $\mathbf{y}$, i.e., such that $\mathbf{T}\mathbf{x}^*=\mathbf{y}$.
In this Section, we provide a general introduction to regularization-based and Bayesian approaches to inverse problems, highlighting the connections between the two. We refer to \cite{stuart_inverse_2010} for an excellent in-depth discussion of these topics.
\subsection{Variational approach: regularization}\label{regularization_approach}
In classical variational (or regularization-based) methods, the solution to the inverse problem is obtained as the minimizer of a suitably defined (unconstrained) penalized optimization problem
\begin{equation}\label{penalized_optimization}
 \mathbf{x}^* = \argmin_{\mathbf{x}\in\mathbb{R}^N}\; L(\mathbf{y}, \mathbf{T}\mathbf{x}) + \lambda R(\mathbf{x})\,,
\end{equation}
where the cost functional $L:\mathbb{R}^M \times\mathbb{R}^M\to\mathbb{R}^+$ enforces data-fidelity, the regularization functional $R:\mathbb{R}^N\to\mathbb{R}^+$ promotes features expected from prior information about the signal being reconstructed, and the regularization parameter $\lambda\in\mathbb{R}^+$ can be tuned to control the amount of regularization.
For the common choice
\begin{equation}\label{least_squares}
    L(\mathbf{y}, \mathbf{T}\mathbf{x}) = \Vert\mathbf{y}- \mathbf{T}\mathbf{x}\Vert_2^2\;,
\end{equation}
Eq.\eqref{penalized_optimization} becomes a penalized least squares problem. The regularization term $R$ is most often chosen as $R(\mathbf{x})=\Vert \mathbf{G}\mathbf{x}\Vert_p^p$ with $p\in\{1,2\}$ for suitable operators $\mathbf{G}$, where, in general, $p=1$ promotes sparsity, $p=2$ smoothness \cite{bertero_introduction_2020, boyd_convex_2004}. The operator $\mathbf{G}$ is typically a discrete differential operator (gradient, Laplacian, $\ldots$) or a transformation decomposing $\mathbf{x}$ into a set of basis functions (wavelets, Fourier, $\ldots$). If the involved functionals and operators satisfy certain properties, the optimization problem in Eq.\eqref{penalized_optimization} is convex, and the existence (and possibly the uniqueness) of a solution is guaranteed \cite{boyd_convex_2004}; such solution can be found applying a suitable optimization algorithm, e.g., (proximal) gradient descent \cite{boyd_convex_2004}.
\subsection{Bayesian approach}\label{bayes_approach}
The choice of the data-fidelity and regularization functionals in Eq.\eqref{penalized_optimization} might appear a bit arbitrary: Bayesian inference provides a rational support for it \cite{stuart_inverse_2010}. In the Bayesian approach, the unknown $\mathbf{x}$ is modelled as a random variable. Given a statistical model of the noise $\boldsymbol{\varepsilon}$ in Eq.\eqref{ip_data}, the likelihood of measuring $\mathbf{y}$ if the ground truth is $\mathbf{x}$ is described by the data \emph{likelihood} probability distribution $p(\mathbf{y}\,\vert\,\mathbf{x})$.
Furthermore, the prior knowledge about $\mathbf{x}$ is assumed to be described by the \emph{prior} distribution $p(\mathbf{x})$. Then, by Bayes' formula, the \emph{posterior} distribution characterising our knowledge about $\mathbf{x}$ given $\mathbf{y}$ is
\begin{equation}\label{bayes_formula}
    p(\mathbf{x}\,\vert \mathbf{y}) = \frac{p(\mathbf{y} \vert \mathbf{x}) p(\mathbf{x})} {p(\mathbf{y})}\;\propto p(\mathbf{y} \vert \mathbf{x}) p(\mathbf{x})\,,
\end{equation}
where the normalizing factor $p(\mathbf{y})=\int_{\mathbb{R}^N}p(\mathbf{y}|\mathbf{x})p(\mathbf{x})\mathrm{d}\mathbf{x}$ is called the \emph{evidence} (or \emph{marginal likelihood}). Inference about $\mathbf{x}$ can now be performed based on the posterior. A popular approach is to compute the Maximum A Posteriori (MAP) estimator, i.e., the maximizer of the posterior in Eq.\eqref{bayes_formula}. It reads\footnote[1]{Since the logarithm is an increasing function, $\argmax_{\mathbf{x}} p(\mathbf{x}\,\vert \mathbf{y})\!=\!\argmax_{\mathbf{x}} \log p(\mathbf{x}\,\vert \mathbf{y})$.}
\begin{equation}\label{MAP}
    \begin{aligned}
        \mathbf{x}_{_{MAP}}& =\argmax_{\mathbf{x}\in\mathbb{R}^N}\; \log p(\mathbf{x}\,\vert \mathbf{y}) \\
        & = \argmax_{\mathbf{x}\in\mathbb{R}^N}\;\log \bigg(\frac{p(\mathbf{y} \vert \mathbf{x}) p(\mathbf{x})}{p(\mathbf{y})}\bigg) \\
        & =\argmax_{\mathbf{x}\in\mathbb{R}^N}\; \log p(\mathbf{y} \vert \mathbf{x}) + \log p(\mathbf{x}) - \log p(\mathbf{y})\\
        & =\argmin_{\mathbf{x}\in\mathbb{R}^N}\; -\log p(\mathbf{y} \vert \mathbf{x}) - \log p(\mathbf{x}),
    \end{aligned}
\end{equation}
where $\log p(\mathbf{y})$ can be neglected since it does not depend on $\mathbf{x}$.
This leads to a key remark: the structures of the problems in Eq.\eqref{MAP} and Eq.\eqref{penalized_optimization} are analogous; however, the ``data-fidelity" and ``regularization" functionals involved in Eq.\eqref{MAP} now have clear statistical meanings.\\In particular, let us consider the situation where the noise $\boldsymbol{\varepsilon}$ in Eq.\eqref{ip_data} can be modelled as additive white Gaussian noise, i.e., independent Gaussian noise of variance $\sigma^2$ for each detector (we could also assume detector-dependent variance $\sigma_m^2$). Then, it follows from Eq.\eqref{ip_data} that the likelihood is Gaussian:
\begin{equation}\label{gaussian_likelihood}
    \mathbf{y}\vert\mathbf{x} \sim \mathcal{N}(\mathbf{T}\mathbf{x}, \sigma^2\mathbf{I})\;\;
    \implies \;\; p(\mathbf{y}\,\vert\,\mathbf{x}) \propto \exp\Big(-\frac{(\mathbf{y}-\mathbf{T}\mathbf{x})^T(\mathbf{y}-\mathbf{T}\mathbf{x})}{2\sigma^2}\Big).
\end{equation}
In this case, the negative log-likelihood in Eq.\eqref{MAP} becomes
\begin{equation}\label{gaussian_loglikelihood}
\begin{aligned}
    -\log p(\mathbf{y}\,\vert\,\mathbf{x}) &= \frac{1}{2\sigma^2}(\mathbf{y}-\mathbf{T}\mathbf{x})^T(\mathbf{y}-\mathbf{T}\mathbf{x})\\
    &= \frac{1}{2\sigma^2} \Vert \mathbf{y}-\mathbf{T}\mathbf{x} \Vert_2^2,
    \end{aligned}
\end{equation}
which coincides with the least squares data-fidelity functional from Eq.\eqref{least_squares} (up to a multiplicative constant). Thus, such a functional naturally arises from the assumption of additive white Gaussian noise. Furthermore, Laplace-/Gaussian-like priors of the form $\exp\big(-\lambda \Vert \mathbf{G}\mathbf{x}\Vert_p^p\big)$ with $p\in\{1,2\}$ lead to the regularization terms $R(\mathbf{x})=\Vert \mathbf{G}\mathbf{x}\Vert_p^p$ discussed in Section \ref{regularization_approach}. The solution to a penalized least squares problem akin to Eq.\eqref{penalized_optimization} can therefore be interpreted as a MAP estimator, where the data-fidelity functional is the negative log-likelihood and the regularization functional is the negative log-prior. This is not the only possible interpretation of penalized likelihood problems, and the choice of $R(\mathbf{x})$ often stems from the empirical observation of its good performance rather than from the belief that $\mathbf{x}$ is actually distributed as the corresponding prior satisfying $-\log p(\mathbf{x})\propto R(\mathbf{x})$ \cite{gribonval_should_2011}. Nevertheless, the described Bayesian interpretation rationally justifies the formulation of the optimization problem in Eq.\eqref{penalized_optimization}, providing a sound statistical framework for probabilistic inference.\\
The MAP estimator is one possible way of extracting information from the posterior, but the Bayesian framework is much richer \cite{stuart_inverse_2010}, naturally enabling the principled treatment of the uncertainty associated to the inverse problem in Eq.\eqref{ip_data}. Rather than providing a single answer to Eq.\eqref{ip_data}, the Bayesian approach describes all the possible answers and their relative \emph{credibility} given our prior knowledge together with a statistical model of the noise. Such a probabilistic, rather than deterministic, approach seems particularly appropriate to deal with complex and ill-posed problems.

\section{Extracting Information from the Posterior}\label{posterior}
The Bayesian approach provides a rich and principled way of statistically describing our knowledge about an unknown quantity given a set of noisy indirect measurements. However, the posterior in Eq.\eqref{bayes_formula} is typically very high-dimensional ($N\gg1$). Unfortunately, extracting information from such a high-dimensional distribution is a notoriously complex task, aside from a few analytically tractable cases (Section \ref{gaussian_prior_section}). In this Section, we describe the (stochastic) gradient flow algorithm \cite{roberts_exponential_1996, durmus_efficient_2018} that we propose for posterior-based inference. This robust and well-understood \cite{durmus_efficient_2018, durmus_nonasymptotic_2017, durmus_high-dimensional_2019} algorithm is general, and can be applied to posteriors arising from a wide class of likelihoods and priors. We dedicate Section \ref{MAP_section} to the deterministic version targeting MAP estimation, Section \ref{MCMC_section} to the stochastic version allowing sampling from the posterior. Throughout these Sections, we assume that the log-posterior is continuously differentiable. However, the discussed algorithms can be adapted to settings featuring non-smooth terms: we address the non-differentiable case at the end of Section \ref{MCMC_section}.
\subsection{MAP estimation: gradient flow}\label{MAP_section}
Let us assume that the negative log-posterior is (strictly) convex \cite{boyd_convex_2004}, which is true for a wide class of important imaging models \cite{durmus_efficient_2018}. In the differentiable case, the MAP from Eq.\eqref{MAP} satisfies the first-order optimality condition
\begin{equation}\label{first_order_opt}
    \nabla \log p(\mathbf{x}_{_{MAP}}\vert\mathbf{y}) = \nabla \log p(\mathbf{y}\vert\mathbf{x}_{_{MAP}}) + \nabla\log p(\mathbf{x}_{_{MAP}}) = 0.
\end{equation}
Additionally, let us assume that $\nabla \log p(\mathbf{x}\vert\mathbf{y})$ is $\beta$-Lipschitz continuous, i.e.,
\begin{equation}\label{lipschitz_continuity}
    \Vert \nabla \log p(\mathbf{x}_1\vert\mathbf{y}) - \nabla \log p(\mathbf{x}_2\vert\mathbf{y}) \Vert_2 \;\leq\;  \beta\Vert \mathbf{x}_1-\mathbf{x}_2 \Vert_2,\quad \forall \;\mathbf{x}_1, \mathbf{x}_2\in\mathbb{R}^N\;;
\end{equation}
when such a $\beta$ exists, it is either analytically known or can be numerically estimated\footnote[2]{For example, in the penalized least squares case $\;\log\,p(\mathbf{x}\vert\mathbf{y})=-\frac{1}{2}\Vert\mathbf{y}-\mathbf{T}\mathbf{x}\Vert_2^2-\frac{\lambda}{2}\Vert\mathbf{G}\mathbf{x}\Vert_2^2\;$, it can be shown that it holds $\beta\leq\Vert\mathbf{T}\Vert^2+\lambda\Vert\mathbf{G}\Vert^2$, with $\Vert\cdot\Vert$ the spectral matrix norm.}. Then, we can compute the MAP by running until convergence the iterative algorithm
\begin{equation}\label{gradient_descent}
   \mathbf{x}_{k+1} = \mathbf{x}_k + \tau \Big( \nabla \log p(\mathbf{y}\vert\mathbf{x}_k) + \nabla\log p(\mathbf{x}_k)\Big),\quad \mathbf{x}_0=\mathbf{x}_0,\quad k\geq0,
\end{equation}
where $\mathbf{x}_0$ is the starting point and $\tau$ is the step-size; convergence is guaranteed if $\tau\leq 1/\beta$ \cite{nesterov_lectures_2018}. Eq.\eqref{gradient_descent} is simply a gradient ascent algorithm on the log-posterior, or, equivalently, gradient descent on the negative log-posterior.
Eq.\eqref{gradient_descent} can be seen as the solution, by discretization in time using an explicit Euler scheme with time-step $\Delta t = \tau$, of the \emph{gradient flow} differential equation
\begin{equation}\label{gradient_flow}
    \frac{\partial{\mathbf{x}}}{\partial{t}} = \nabla \log p(\mathbf{y}\vert\mathbf{x}) + \nabla\log p(\mathbf{x}),\quad \mathbf{x}(0)=\mathbf{x}_0,
\end{equation}
where $t$ represents an artificial time. Therefore, the MAP estimate can be seen as the steady-state solution of the partial differential equation (PDE) in Eq.\eqref{gradient_flow}, since the MAP satisfies Eq.\eqref{first_order_opt}.\\More efficient algorithms than the first-order gradient descent in Eq.\eqref{gradient_descent} might (and often do) exist. However, the algorithm in Eq.\eqref{gradient_descent} is robust and each iteration is typically computationally cheap; moreover, in this work, our interest in Eqs.\eqref{gradient_descent}-\eqref{gradient_flow} is mainly due to their stochastic versions, which we introduce next.   

\subsection{Posterior sampling: stochastic gradient flow}\label{MCMC_section}
We attempt to provide a solid intuition of how the stochastic gradient flow algorithm works, whilst keeping the discussion as mathematically light as is possible, by referring to relevant references for further details.\\
As discussed in Section \ref{bayes_approach}, the Bayesian framework is not limited to MAP computation.
\emph{Sampling} provides a way to extract richer information from the posterior distribution. By generating samples $\{\mathbf{x}_k\}_{k=1}^K$ which are distributed according to the posterior, we can perform posterior-based uncertainty quantification and other kinds of advanced statistical analysis.
Markov Chain Monte Carlo (MCMC) techniques \cite{kroese_handbook_2011} are often used for this. Such techniques allow approximate sampling from arbitrary probability distributions.
MCMC methods generate a Markov chain \cite{kroese_handbook_2011} of approximate samples of a target distribution; at each step of the chain, a \emph{proposal distribution} provides a candidate for the next sample. Typically, the target distribution is only the \emph{limiting distribution} of the chain: therefore, the first $K_{b}$ samples are discarded in a so-called \emph{burn-in} phase, since their distribution is significantly different from the limiting one. \emph{Thinning}, i.e., retaining only every $k$-th sample, can additionally be used to reduce correlation between samples.\\
Sampling from high-dimensional distributions, such as those arising from computational imaging inverse problems, is typically computationally intensive, since the efficiency of MCMC techniques deteriorates as the dimension, $N$, increases (see \cite{pereyra_survey_2015} for a review).
However, sampling efficiency in high dimensions can be significantly improved \cite{pereyra_survey_2015} by employing a class of advanced MCMC algorithms featuring a proposal distribution which originates from the discretization of the (overdamped) \emph{Langevin} stochastic differential equation (SDE) \cite{pavliotis_stochastic_2014}
\begin{equation}\label{langevin_sde}
    \mathrm{d}\mathbf{x}_t= \Big(\nabla \log p(\mathbf{y}\vert\mathbf{x}_t) + \nabla\log p(\mathbf{x}_t)\Big)\mathrm{d}t + \sqrt{2}\;\mathrm{d}W_t\;.
\end{equation}
Here, $(W_t)_{t\geq0}$ is the standard $N$-dimensional Brownian motion \cite{pavliotis_stochastic_2014}, used to model randomness. SDEs \cite{pavliotis_stochastic_2014, evans_introduction_2006} are the stochastic version of PDEs: rather than functions, their solutions are stochastic processes, i.e., collections of random variables $(\mathbf{x}_t)_{t\geq0}$ indexed by a variable $t$, typically representing time. For a given $t$, $\mathbf{x}_t$ is a random variable distributed as $\pi_t$, with density $\pi_t(\mathbf{x})$. 
The solution $\mathbf{x}_t$ to a SDE is said to be ergodic if it admits a unique \emph{invariant} (or \emph{stationary}) distribution $\pi_\infty$, i.e., a distribution such that if $\mathbf{x}_{t^*}$ is distributed as $\pi_\infty$ for some $t^*\geq0$, then the distribution of $\mathbf{x}_t$ will be equal to $\pi_{\infty}$ $\forall\, t\geq t^*$. Under mild assumptions, the Langevin SDE in Eq.\eqref{langevin_sde} admits the posterior $p(\mathbf{x}\vert\mathbf{y})$ as unique invariant distribution \cite{durmus_nonasymptotic_2017, pavliotis_stochastic_2014}, and the distribution of $\mathbf{x}_t$ converges to it for $t\to\infty$ for any initial condition $\mathbf{x}_0$ \cite{roberts_exponential_1996, pavliotis_stochastic_2014}.
The invariant distribution is the stochastic equivalent of the steady-state solution of a PDE.
Crucially, ergodic processes satisfy, for any continuous function $\phi:\mathbb{R}^N\to\mathbb{R}$, the \emph{ergodic theorem} \cite{pavliotis_stochastic_2014}
\begin{equation}\label{ergodic_theorem}
    \lim_{T\to\infty}\frac{1}{T}\int_0^T\phi(\mathbf{x}_t)\mathrm{d}t = \int_{\mathbb{R}^N}\phi(\mathbf{x})\pi_\infty(\mathbf{x})\mathrm{d}\mathbf{x},
\end{equation}
which states that time-averaging on long time scales is equivalent to computing expectations with respect to the stationary distribution $\pi_\infty$ (ensemble-averaging).\\
The SDE in Eq.\eqref{langevin_sde} features a deterministic \emph{drift} component and a stochastic \emph{diffusion} term.
The deterministic part, analogous to Eq.\eqref{gradient_flow}, drives $\mathbf{x}$ towards regions of high posterior density. The stochastic part, instead, introduces randomness and allows to explore other areas of the distribution (instead of just the MAP). The \emph{stochastic gradient flow} in Eq.\eqref{langevin_sde} is simply the stochastic version of the gradient flow in Eq.\eqref{gradient_flow}.\\
Since $\pi_\infty(\mathbf{x})=p(\mathbf{x}\,\vert\,\mathbf{y})$, the distribution of $\mathbf{x}_t$ stops evolving in time once it has converged to the posterior distribution. This key property enables efficient sampling from the posterior.
Samples can, for instance, be acquired by the \emph{Unadjusted Langevin Algorithm} (ULA) \cite{roberts_exponential_1996, durmus_efficient_2018}
\begin{equation}\label{ULA}
   \mathbf{x}_{k+1} = \mathbf{x}_k + \tau \Big( \nabla \log p(\mathbf{y}\vert\mathbf{x}_k) + \nabla\log p(\mathbf{x}_k)\Big) + \sqrt{2\tau}\,\mathbf{z}_k,\quad k\geq0,
\end{equation}
where $(\mathbf{z}_k)_{k\geq0}\sim\mathcal{N}(\mathbf{0},\mathbf{I})$ are independent realizations of standard $N$-dimen-sional Gaussian vectors.
The discrete-time Markov chain in Eq.\eqref{ULA} is the explicit Euler-Maruyama discretization \cite{pavliotis_stochastic_2014} with time step $\Delta t =\tau$ of Eq.\eqref{langevin_sde}. This is the stochastic analogous of the explicit Euler method highlighted for Eqs.\eqref{gradient_descent}-\eqref{gradient_flow}. For log-posteriors satisfying Eq.\eqref{lipschitz_continuity}, the condition $\tau\leq 1/\beta$ guarantees stability of the chain in Eq.\eqref{ULA} and its convergence to a unique stationary distribution $\pi_{\infty\,;\,\tau}\,$. Typically, $\pi_{\infty\,;\,\tau}$ is different from $\pi_\infty(\mathbf{x})=p(\mathbf{x}\vert\mathbf{y})$, but converges to it for $\tau\to0$ \cite{durmus_nonasymptotic_2017}, i.e., for vanishing discretization error. The \emph{bias} introduced by such difference can be controlled by reducing $\tau$. In practice, for high-dimensional problems, values close to the upper bound $1/\beta$ are selected, to favor efficiency at the expense of some (typically small) bias \cite{durmus_efficient_2018}.
While other variants of ULA have traditionally been preferred \cite{roberts_exponential_1996}, recent convergence results \cite{durmus_nonasymptotic_2017, durmus_high-dimensional_2019} have contributed to a resurgence of interest in Eq.\eqref{ULA} for efficient high-dimensional sampling \cite{durmus_efficient_2018}.\\
Let us assume that we generate $K_{tot} \!=\! K_b \!+\! K \!+\!1$ samples $\{\mathbf{x}_k\}_{k=0}^{K_{tot}}$ from Eq.\eqref{ULA}, simulating it up to the time $T_{tot}=K_{tot}\tau$. By the ergodic theorem in Eq.\eqref{ergodic_theorem}, defining $T\!=\!(K+1)\tau$, we can approximate expectations of functionals of interest $\phi$ with respect to the posterior by the Monte Carlo estimate
\begin{equation}\label{MC_estimate}
\begin{aligned}
\int_{\mathbb{R}^N}\phi(\mathbf{x})p(\mathbf{x}\vert\mathbf{y})\mathrm{d}\mathbf{x} &\approx \frac{1}{T} \sum_{k=K_b}^{K_{tot}} \tau \phi(\mathbf{x}_k)\\
&= \frac{1}{K+1} \sum_{k=K_b}^{K_{tot}} \phi(\mathbf{x}_k),
\end{aligned}
\end{equation}
with $K_b$ the number of burn-in iterations. We can, for example, use the samples to estimate moments of the posterior (mean for $\phi(x)=x$, variance for $\phi(x)=x^2$, skewness, kurtosis, $\ldots$).
Crucially, such estimates can be computed using an \emph{online algorithm} \cite{welford_note_1962}. It is thus not necessary to store all the samples $\{\mathbf{x}_k\}_{k=K_b}^{K_{tot}}$ for MC estimation by Eq.\eqref{MC_estimate}, which would be very memory intensive. Instead, at each ULA step, we simply discard the acquired sample after it has been used to update the MC estimates.

\begin{remark}\label{remark_differentiable}
In Sections \ref{MAP_section} and \ref{MCMC_section}, we focused on continuously differentiable log-posteriors. However, this requirement can be weakened. This is important, as several commonly used priors are not continuously differentiable. Important examples are the sparsity promoting priors $\propto \exp(-\lambda\Vert\mathbf{G}\mathbf{x}\Vert_1)$, with $\mathbf{G}$ introduced in Section \ref{regularization_approach}, or the prior $\mathcal{P}_{\mathbf{x}\geq\boldsymbol{0}}=\exp(-\mathbbm{1}_{\mathbf{x}\geq\boldsymbol{0}})$ associated to the positivity constraint $\mathbbm{1}_{\mathbf{x}\geq\boldsymbol{0}}$.
\emph{Proximal gradient descent} algorithms \cite{combettes_proximal_2011, parikh_proximal_2014} for MAP computation and \emph{proximal MCMC} methods \cite{durmus_efficient_2018, cai_uncertainty_2018-1} for posterior sampling generalize the discussed techniques to a larger class of posteriors; we describe proximal techniques in Appendix \ref{appendix_prox}.
\end{remark}

\section{Plasma Tomography: Towards a Unifying View}\label{unifying}
As discussed in Section \ref{introduction}, tomography is used to reconstruct the spatial emissivity distribution in fusion devices such as tokamaks. Typically, rather than attempting a 3D reconstruction of the emissivity over the entire 3D volume, tomographic techniques are used in tokamaks to estimate the 2D emissivity pattern on a poloidal cross-section, perpendicular to the toroidal direction. This is achieved by designing diagnostics that cover a specific poloidal cross-section, and/or by assuming that the signal to be reconstructed is toroidally symmetric \cite{ingesson_chapter_2008}. Under the latter assumption, the reconstructed 2D emissivity can then be used to estimate the total radiated power over a given energy range.\\
Throughout this paper, we focus on tomography diagnostics covering a single poloidal cross-section $\Omega$. Moreover, we expand the emissivity function in a set of $N=N_z\times N_r$ rectangular pixels, with $N_z$ and $N_r$ the number of pixels used to discretize $\Omega$ in the vertical and radial directions, respectively.
With this choice of basis functions\footnote[3]{Other types of basis functions could be chosen \cite{ingesson_chapter_2008}, for example splines \cite{vezinet_non-monotonic_2016}.}, let $\mathbf{x}\in\mathbb{R}^N$ be the plasma emissivity poloidal distribution. Then, let $\mathbf{T}\in\mathbb{R}^{M\times N}$ model the acquisition of tomographic data for a diagnostic composed of $M$ detectors. $\mathbf{T}$ is typically modeled as a linear operator referred to as geometry (or sensitivity) matrix \cite{ingesson_chapter_2008, jardin_use_2017}, with $T_{ij}$ characterizing the contribution of the $j$-th pixel to the measurement of the $i$-th detector. The observed noisy tomographic data are described by Eq.\eqref{ip_data}.
As discussed in Section \ref{introduction}, many techniques have been applied to tomographic reconstruction from fusion devices' data. In this Section, we show that many of them fit within the general Bayesian framework described in Section \ref{bayes_approach}. Their shared common traits uncover deep and interesting connections between the approaches, promoting a unifying view. In Section \ref{reaction_diffusion_section}, we consider gradient-penalizing smoothness priors, thus introducing the prior class that we rely upon in Section \ref{application}. Then, in Section \ref{gaussian_prior_section}, we discuss the generalization of Gaussian Process Tomography to a wide class of popular smoothness priors, which includes some of the priors discussed in Section \ref{reaction_diffusion_section}. 
\begin{remark}\label{remark_noise_model}
As discussed in Section \ref{bayes_approach}, the likelihood depends on the model chosen to describe the noise $\boldsymbol{\varepsilon}$ in Eq.\eqref{ip_data}. Throughout this work, we will mainly consider additive white Gaussian noise such that $\varepsilon_m\sim\mathcal{N}(0, \sigma_m^2),\;m\!=\!1,\ldots,M$, leading to a Gaussian likelihood. Moreover, in Sections \ref{reaction_diffusion_section} and \ref{gaussian_prior_section}, we assume, for simplicity, $\sigma_m=\sigma,\,m\!=\!1,\ldots,M$. Here, as discussed in Section \ref{bayes_approach}, the negative log-likelihood is given by Eq.\eqref{gaussian_loglikelihood}. However, other noise models could be employed. For example, we could consider a Poisson noise model: the resulting likelihood \cite{bertero_introduction_2020} can be combined with priors akin to those discussed in Section \ref{bayes_approach}, thus defining a posterior-based inference framework. A Poisson model is typically assumed in the context of Maximum Likelihood (ML) tomography \cite{craciunescu_maximum_2008, craciunescu_evaluation_2016, craciunescu_maximum_2018}, which achieves regularization by means of filtering techniques and implements a non-Bayesian UQ technique. A comparative study between the ML method and a fully Bayesian approach based on the inference techniques described in Section \ref{posterior} would be of great interest but remains beyond the scope of this work.
\end{remark}
\begin{remark}\label{remark_constrained_vs_unconstrained}
The Bayesian approach naturally leads to unconstrained optimization problems in the form of Eq.\eqref{penalized_optimization} (or, equivalently, Eq.\eqref{MAP}). However, several authors in the plasma tomography community prefer to consider the constrained optimization version of such problem, which reads
\begin{equation}\label{constrained_optimization}
    \text{minimize} \; R(\mathbf{x}),\quad\text{subject to} \;\;\chi^2=\frac{1}{\sigma^2}\Vert\mathbf{y}-\mathbf{T}\mathbf{x}\Vert_2^2\leq M,
\end{equation}
where the constraint controls the difference between the observed and reconstructed measurements \cite{ingesson_chapter_2008, anton_x-ray_1996}. In practice, solving Eq.\eqref{constrained_optimization} typically requires solving a sequence of problems similar to the unconstrained Eq.\eqref{penalized_optimization}, where the regularization parameter $\lambda$ is iteratively updated to find a value such that the constraint in Eq.\eqref{constrained_optimization} is satisfied. As discussed in \cite{ingesson_chapter_2008}, the two formulations are closely related. In this work, since we are interested in the Bayesian setting, we will only consider unconstrained formulations.
\end{remark}
\subsection{Gradient-penalizing smoothness priors: reaction-diffusion imaging}\label{reaction_diffusion_section}
Many smoothing techniques proposed for plasma tomography rely on quadratic \emph{generalized Tikhonov} regularization functionals
\begin{equation}\label{generalized_tikhonov}
R(\mathbf{x}) = \frac{1}{2}\Vert\mathbf{G}\mathbf{x}\Vert_2^2 = \frac{1}{2}\mathbf{x}^T\mathbf{G}^T\mathbf{G}\mathbf{x},
\end{equation}
where the matrix $\mathbf{G}$ is related to discretized zeroth/first/second order derivative operators \cite{ingesson_chapter_2008}, typically computed via finite differences. In this Section, we focus on approaches where $\mathbf{G}\in\mathbb{R}^{2N\times N}$ depends on the discrete gradient operator $\boldsymbol{\nabla}=(\boldsymbol{\nabla}_z; \boldsymbol{\nabla}_r)\in\mathbb{R}^{2N\times N}$. Such approaches aim at penalizing strong gradients in the reconstruction, optionally leveraging knowledge of the magnetic equilibrium and/or introducing MFI-inspired weighting terms.
Choosing $\mathbf{G}^T\mathbf{G}=\boldsymbol{\nabla}^T\boldsymbol{\nabla}$ isotropically penalizes the gradient, while $\mathbf{G}^T\mathbf{G}=\alpha_\parallel\boldsymbol{\nabla}_{\parallel}^T\boldsymbol{\nabla}_{\parallel}+\alpha_\perp\boldsymbol{\nabla}_{\perp}^T\boldsymbol{\nabla}_{\perp}$ achieves anisotropic regularization, typically used to favor smoothness along flux surfaces by imposing $\alpha_\perp<\alpha_\parallel$ \cite{ingesson_chapter_2008}. In MFI regularization \cite{anton_x-ray_1996}, a diagonal weight matrix $\mathbf{W}(\mathbf{x})$
is introduced to increase (resp. decrease) smoothing in regions of low (resp. high) emissivity. For isotropic MFI, $\mathbf{G}^T\mathbf{G}=\big(\mathbf{G}^T\mathbf{G}\big)(\mathbf{x})=\boldsymbol{\nabla}_{z}^T\mathbf{W}(\mathbf{x})\boldsymbol{\nabla}_{z}+\boldsymbol{\nabla}_{r}^T\mathbf{W}(\mathbf{x})\boldsymbol{\nabla}_{r}$; the anisotropic version is similarly defined. In all such cases, we can rewrite $\mathbf{G}^T\mathbf{G}$ as $\boldsymbol{\nabla}^T(\mathbf{D}\boldsymbol{\nabla})$, where $\mathbf{D}\in\mathbb{R}^{2N\times 2N}$ is a tensor defining the local smoothing properties (direction and strength). $\mathbf{D}$ associates to each pixel $n$ a $2\times2$ positive semidefinite matrix $\mathbf{D}_n$ acting on the local gradient:
\begin{equation}\label{eq_diff_coeff}
\begin{aligned}
& \mathbf{D}_n=\mathbf{I} \qquad\qquad\qquad\qquad\qquad\;\,\mbox{for isotropic smoothing};\\
&\mathbf{D}_n=\alpha_{\parallel}\mathbf{e}_{n,\parallel}\mathbf{e}_{n,\parallel}^{T}+\alpha_{\perp}\mathbf{e}_{n,\perp}\mathbf{e}_{n,\perp}^{T}\;\; \mbox{for anisotropic smoothing, with}\\
& \qquad\qquad\qquad\qquad\quad\qquad\qquad\;\; \mathbf{e}_{n,\parallel}, \mathbf{e}_{n,\perp} \;\mbox{the unit vectors locally}\\
& \qquad\qquad\qquad\qquad\quad\qquad\qquad\;\; \mbox{pointing along/across flux surfaces;}\\
&\mathbf{D}_n=W_{nn}(x_n)\;\mathbf{I} \qquad\qquad\qquad\;\,\mbox{in the isotropic MFI case, with}\\
&\qquad\qquad\qquad\qquad\quad\qquad\qquad\;\; W_{nn}(x_n)=1/\max\{\delta,\,x_n\},\;\delta\ll1.
\end{aligned}
\end{equation} 
Then\footnote[4]{Considering, in the MFI case, its usual linearized version \cite{anton_x-ray_1996}, thus neglecting the extra term arising from differentiation of $W(\mathbf{x})$.}, the gradient of the quadratic form Eq.\eqref{generalized_tikhonov} reads $\nabla R(\mathbf{x})=\boldsymbol{\nabla}^T\!(\mathbf{D}\boldsymbol{\nabla}\mathbf{x})$. 
As the divergence is the negative adjoint of the gradient, it holds that $\boldsymbol{\nabla}^T(\mathbf{D}\boldsymbol{\nabla}\mathbf{x})=-\boldsymbol{\nabla}\cdot(\mathbf{D}\boldsymbol{\nabla}\mathbf{x})$. The tensor $\mathbf{D}$ can thus be interpreted as a \emph{diffusion tensor}, which can be designed to obtain a regularization term promoting desired features.
The diffusion analogy of isotropic/anisotropic regularization is well-known to the plasma community in a slightly different form \cite{ingesson_chapter_2008, fuchs_2d_1994, ingesson_soft_1998}. However, such analogy becomes even clearer with our gradient-based definition of the regularization functional. Indeed, let us consider the gradient flow Eq.\eqref{gradient_flow} towards the MAP: under the noise modelling assumptions of Remark \ref{remark_noise_model}, such a flow becomes the reaction-diffusion \cite{weickert_anisotropic_1998} equation
\begin{equation}\label{reaction-diffusion}
    \frac{\partial{\mathbf{x}}}{\partial{t}} =\underbrace{\frac{1}{\sigma^2}\mathbf{T}^T(\mathbf{y}-\mathbf{T}\mathbf{x})}_\text{reaction term} + \lambda \underbrace{\!\!\!\phantom{\frac{1}{2}}\!\!\boldsymbol{\nabla}\cdot(\mathbf{D}\boldsymbol{\nabla}\mathbf{x})}_\text{diffusion term},
\end{equation}
where the likelihood contribution can be viewed as a reaction term acting as an emissivity source/sink to ensure data-fidelity. We emphasize that, as discussed in \cite{ingesson_chapter_2008, ingesson_mathematics_2000}, the diffusion in Eq.\eqref{reaction-diffusion} should not be interpreted as a real physics plasma transport process: it simply models a mathematical diffusion of emissivity values.\\
The (reaction-)diffusion analogy has been extensively investigated by the image processing community. In particular, PDE-based image smoothing methods have been proposed, with the smoothing achieved by computing the time evolution of the \emph{diffusion equation} $\partial{\mathbf{x}}/\partial{t} = \boldsymbol{\nabla}\cdot(\mathbf{D}(\mathbf{x})\boldsymbol{\nabla}\mathbf{x})$;
 we refer to \cite{weickert_anisotropic_1998} for an excellent review on this rich topic. For $\mathbf{D}=\mathbf{I}$, the diffusion term in Eq.\eqref{reaction-diffusion} is simply $\boldsymbol{\Delta}\mathbf{x}$, leading to isotropic Gaussian/Laplacian smoothing \cite{weickert_anisotropic_1998}.
More sophisticated isotropic/anisotropic tensors, that allow, for example, edge-enhancement, coherence-enhancement, and/or curvature-preservation, have also been proposed \cite{perona_scale-space_1990, weickert_anisotropic_1998, tschumperle_vector-valued_2005, tschumperle_anisotropic_2007}.
 Finally, the idea of introducing a reaction term for image reconstruction purposes is well-known too \cite{nordstrom_biased_1990, rudin_nonlinear_1992, weickert_anisotropic_1998}. PDE-based smoothing, based on \cite{tschumperle_anisotropic_2007}, was recently proposed \cite{craciunescu_maximum_2023} in the context of Maximum Likelihood-based plasma emission tomography.\\
 In this work, we always consider diffusion terms such that $R(\mathbf{x})$ is well-defined and given by Eq.\eqref{generalized_tikhonov} for $\mathbf{G}^T\mathbf{G}=\boldsymbol{\nabla}^T\mathbf{D}\boldsymbol{\nabla}$, i.e., $\mathbf{G}=\sqrt{\mathbf{D}}\boldsymbol{\nabla}$ where by $\sqrt{\mathbf{D}}$ we denote the unique positive semidefinite matrix satisying $\sqrt{\mathbf{D}}\sqrt{\mathbf{D}}=\mathbf{D}$. The operator $\boldsymbol{\nabla}$ is computed through finite differences.
  Then, as discussed in Sections \ref{bayes_approach} and \ref{MAP_section}, all such regularization terms admit a Bayesian interpretation: we can interpret the steady-state of Eq.\eqref{reaction-diffusion} as a MAP estimator, with prior $p(\mathbf{x})\propto\exp\big(-\lambda R(\mathbf{x})\big)$. Moreover, posterior sampling, by means of ULA (Eq.\eqref{ULA}), can be performed. All the plasma tomography-related regularization techniques introduced at the beginning of this Section fit within a Bayesian framework.

\subsection{Gaussian smoothness priors and Gaussian Process Tomography}\label{gaussian_prior_section}
Methods based on Gaussian processes have emerged as a popular way of performing Bayesian inference; we suggest \cite{rasmussen_gaussian_2005} for a general introduction to the topic. In the context of plasma imaging, this led to the development of Gaussian Process Tomography approaches \cite{svensson_nonparametric_2011, li_bayesian_2013, wang_incorporating_2018, wang_gaussian_2018, moser_gaussian_2022}. Here, both the likelihood and the prior are modeled as Gaussian distributions. The likelihood, as discussed in Remark \ref{remark_noise_model}, is given by Eq.\eqref{gaussian_likelihood}, and the prior by
\begin{equation}\label{gaussian_process_tomograhy_prior}
\begin{aligned}
&\mathbf{x}\sim\mathcal{N}(\mathbf{0},\,\boldsymbol{\Sigma})\\
&p(\mathbf{x})\propto \exp\Big(-\frac{1}{2}\mathbf{x}^T\boldsymbol{\Sigma}^{-1}\mathbf{x}\Big),
\end{aligned}
\end{equation}
where the prior covariance matrix $\boldsymbol{\Sigma}\in\mathbb{R}^{N\times N}$ is typically defined as \cite{svensson_nonparametric_2011, li_bayesian_2013, wang_gaussian_2018}
\begin{equation}\label{kernel_function}
    \boldsymbol{\Sigma}_{n_1,n_2} = k\Big((z_{n_1}, r_{n_1}), (z_{n_2}, r_{n_2})\Big),
\end{equation}
where the kernel function $k$ describes the correlation between pixels as a function of their poloidal locations $(z_{n_i},r_{n_i})$; kernel functions allowing the inclusion of magnetic equilibrium information have also been proposed \cite{wang_incorporating_2018, moser_gaussian_2022}. The key feature of Gaussian Process Tomography is the Gaussianity of the resulting posterior, which is given by $\mathbf{x}\vert\mathbf{y}\sim\mathcal{N}(\boldsymbol{\mu}_{post},\,\boldsymbol{\Sigma}_{post})$, with
\begin{equation}\label{gaussian_mean_cov}
    \boldsymbol{\mu}_{post} = \sigma^{-2}\,\boldsymbol{\Sigma}_{post} \mathbf{T}^T\mathbf{y},\quad \boldsymbol{\Sigma}_{post} = \big( \boldsymbol{\Sigma}^{-1}\,+\,\sigma^{-2}\,\mathbf{T}^T\mathbf{T} \big)^{-1}.
\end{equation}
Therefore, the mean and the covariance matrix of the posterior distribution are available in closed-form, allowing efficient MAP computation and UQ. Here, $\boldsymbol{\mu}_{post}$ coincides with the MAP as the distribution is Gaussian, while $\boldsymbol{\Sigma}_{post}$ provides an estimate of the uncertainty in the reconstruction for the given noise model and prior's assumptions.\\
In Gaussian Process Tomography, the prior is defined through a kernel function. Gaussian posteriors, however, arise from a much wider class of priors. Indeed, let us consider the class of \emph{Gaussian smoothness priors} \cite{kaipio_statistical_2005}
\begin{equation}\label{gaussian_smothness_prior}
p(\mathbf{x})\propto \exp\Big(-\frac{1}{2}\,\Vert\mathbf{G}\mathbf{x}\Vert_2^2\Big) = \exp\Big(-\frac{1}{2}\,\mathbf{x}^T\mathbf{G}^T\mathbf{G}\mathbf{x}\Big),
\end{equation}
for matrices $\mathbf{G}$ which do not depend on $\mathbf{x}$; matrices $\mathbf{G}=\mathbf{G}(\mathbf{x})$, such as those arising in MFI-based techniques, cannot be expected to lead to Gaussian posteriors.
As discussed in Section \ref{reaction_diffusion_section}, we are especially interested in smoothness priors with $\mathbf{G}$ defined as a function of a discrete differential operator. Depending on the prescribed boundary conditions, the matrix $\mathbf{G}$ may or may not be of full rank. When rank-deficient, the resulting prior is \emph{improper}: $p(\mathbf{x})$ cannot be interpreted as a proper probability density, because it is not integrable \cite{kaipio_statistical_2005}. Therefore, technically, the prior is not Gaussian, despite the similarity of the improper density Eq.\eqref{gaussian_smothness_prior} with Eq.\eqref{gaussian_process_tomograhy_prior}. In particular, with a rank-deficient $\mathbf{G}$, $\mathbf{G}^T\mathbf{G}$ is not invertible: as a consequence, there is no covariance matrix $\boldsymbol{\Sigma}_{prior}$ such that $\boldsymbol{\Sigma}_{prior}^{-1}=\mathbf{G}^T\mathbf{G}$ in Eq.\eqref{gaussian_smothness_prior}. Nevertheless, by \cite[Theorem 3.10]{kaipio_statistical_2005}, it holds that the resulting posterior is proper and Gaussian even for improper priors so long as $\mathrm{Ker}(\mathbf{T})\cap \mathrm{Ker}(\mathbf{G})=\{\mathbf{0}\}$, i.e., if the null spaces of the matrices $\mathbf{T}$ and $\mathbf{G}$ only share the null vector. When this condition is met, $\mathbf{x}\vert\mathbf{y}\sim\mathcal{N}(\boldsymbol{\mu}_{post}, \boldsymbol{\Sigma}_{post})$, where the posterior mean and covariance matrix are obtained by replacing $\boldsymbol{\Sigma}^{-1}$ by $\mathbf{G}^T\mathbf{G}$ in Eq.\eqref{gaussian_mean_cov}. These formulas also hold, of course, when the prior is proper.\\
Many smoothness priors used for plasma tomography satisfy such a condition. 
The Tikhonov prior ($\mathbf{G}^T\mathbf{G}=\mathbf{I}$) is proper. Gradient-based priors such that $\mathbf{G}^T\mathbf{G}=\boldsymbol{\nabla}^T\mathbf{D}\boldsymbol{\nabla}$ can be proper, or improper, depending on the boundary conditions; for no-flux boundary conditions, for example, it is improper, but $\mathrm{Ker}(\mathbf{G})$ only contains constant vectors which do not belong to $\mathrm{Ker}(\mathbf{T})$ (apart from the trivial null vector). Similar arguments hold for commonly used \cite{ingesson_chapter_2008} operators $\mathbf{G}$ involving second order differentiation operators. An even stronger statement can thus be made upon the Bayesian interpretation, discussed at the end of Section \ref{reaction_diffusion_section}, of many plasma tomography approaches: some of the most popular priors lead to Gaussian posteriors, for which the MAP and UQ are readily available. Gaussian Process Tomography can be extended to the wide class of Gaussian smoothness priors. In particular, for $\mathbf{G}^T\mathbf{G}=\boldsymbol{\nabla}^T\mathbf{D}\boldsymbol{\nabla}$, the choice of diffusion coefficient replaces the choice of kernel function in Eq.\eqref{kernel_function}. 

\section{Application: SXR Imaging at TCV}\label{application}
\begin{figure}[]{}
\includegraphics[width=\textwidth]{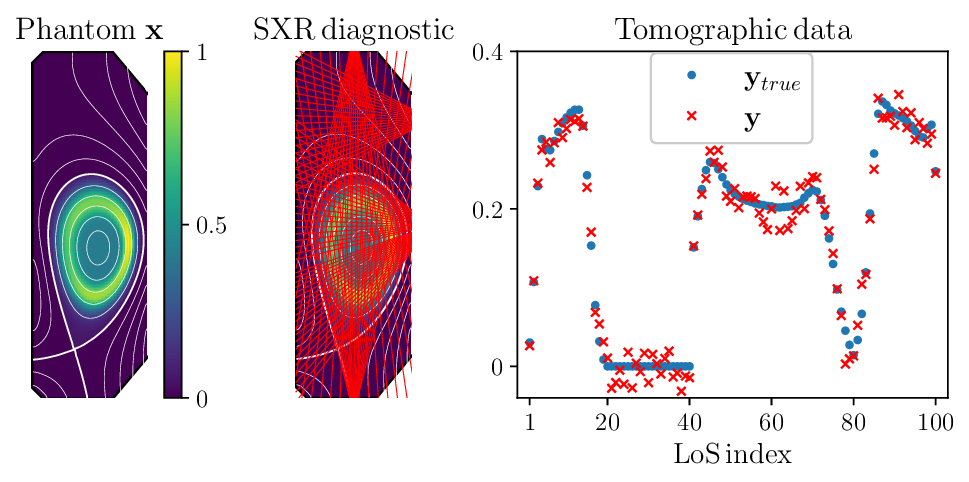}
\vspace{-0.8cm}
\captionsetup{justification=raggedright,singlelinecheck=true}
\caption{SXR diode system installed at TCV. From left to right: poloidal cross-section of TCV showing a SXR phantom model $\mathbf{x}$, with magnetic flux surfaces in white; line of sight (LoS) configuration overlaid to phantom $\mathbf{x}$; corresponding tomographic data, without ($\mathbf{y}_{true}$) and with ($\mathbf{y}$) noise added.}
\label{fig_diag}
\end{figure}
To illustrate the main features of the imaging framework proposed in this paper, we apply it to SXR imaging at TCV.
SXR tomography of fusion plasmas is a long-established tool which can be used to study MHD activity, impurity transport and other phenomena \cite{anton_x-ray_1996, ingesson_soft_1998, ingesson_chapter_2008}. SXR diagnostics are installed at most tokamak facilities. At TCV, SXR imaging relies on a silicon photodiode system now integrated as a part of the RADCAM diagnostic \cite{sheikh_radcam-radiation_2022}. The system is composed of $100$ diodes, organized within $5$ cameras of $20$ diodes. In Figure \ref{fig_diag}, we provide a visual representation of how the SXR diodes installed at TCV probe plasma emissivity. The noiseless tomographic measurements $\mathbf{y}_{true}$ correspond, to a first approximation, to the integrals of the emissivity along the diodes' lines of sight\footnote[5]{Up to a scaling factor (etendue) modeling the geometrical properties of the system \cite{ingesson_chapter_2008}.};
in practice, the observed measurements $\mathbf{y}$ are a noise-corrupted  version of $\mathbf{y}_{true}\,$, as expressed in Eq.\eqref{ip_data}.\\
To assess the quality of any plasma tomography algorithm, it is common practice to test its performance on model phantoms, i.e., synthetic emissivity profiles. This allows directly comparing the algorithms' output to a ground truth, which cannot be done when working with real data. Ideally, algorithms should be tested on a phantom set as large, varied and physically realistic as possible. In Section \ref{dataset}, we describe the generation of a large dataset of $10^3$ diverse phantoms. Then, in Section \ref{results}, we apply the techniques described in Sections \ref{MAP_section}--\ref{MCMC_section} to this dataset, demonstrating how posterior-based inference enables insightful statistical treatments of the tomographic problem. Finally, in Section \ref{limitations}, we discuss the inherent limitations of plasma sparse-view tomography, rooted in the extremely ill-posed nature of this inverse problem.\\
All the data, results and analysis presented in this paper can be reproduced with the open access code provided at the dedicated GitHub repository \url{https://github.com/dhamm97/bayes-plasma-tomo.git}. For computations, we rely on \texttt{Pyxu} \cite{pyxu-framework}, an open-source computational imaging Python framework developed, among others, by the authors reported as affiliated to the EPFL Center for Imaging.
Due to its module agnostic codebase, computations can be performed on CPUs or GPUs. The plasma imaging regularization terms discussed in this work are implemented by the \texttt{Pyxu} plug-in \texttt{pyxu-diffops} \cite{pyxu-diffops}. All computations were performed on a workstation with Intel Core i9-10900X processor (10 CPU cores) and 128GB of RAM.

\subsection{Dataset generation}\label{dataset}\begin{figure}[]
\begin{subfigure}[b]{0.14\textwidth}
         \centering
         \includegraphics[width=\textwidth]{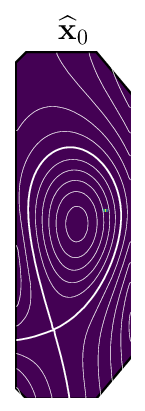}
         \caption{}
         \label{fig_phantom_gen0}
     \end{subfigure}
     \hspace{-0.4cm}
\begin{subfigure}[b]{0.14\textwidth}
         \centering
         \includegraphics[width=\textwidth]{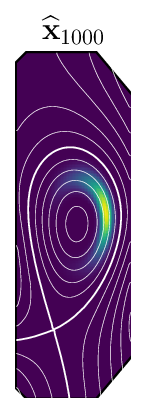}
         \caption{}
         \label{fig_phantom_gen1}
     \end{subfigure}
     \hspace{-0.4cm}
\begin{subfigure}[b]{0.14\textwidth}
         \centering
         \includegraphics[width=\textwidth]{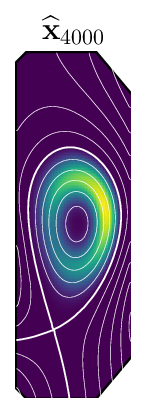}
         \caption{}
         \label{fig_phantom_gen2}
     \end{subfigure}
     \hspace{-0.4cm}
\begin{subfigure}[b]{0.14\textwidth}
         \centering
         \includegraphics[width=\textwidth]{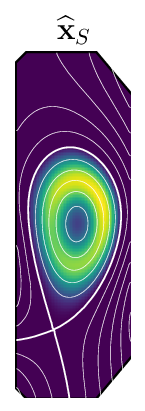}
         \caption{}
         \label{fig_phantom_gen3}
     \end{subfigure}
     \hspace{0.4cm}
\begin{subfigure}[b]{0.14\textwidth}
         \centering
         \includegraphics[width=\textwidth]{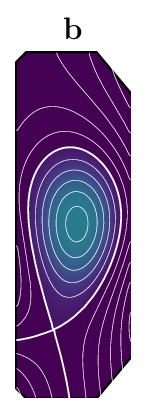}
         \caption{}
         \label{fig_phantom_gen_b}
     \end{subfigure}
     \hspace{0.2cm}
\begin{subfigure}[b]{0.1725\textwidth}
         \centering
         \includegraphics[width=\textwidth]{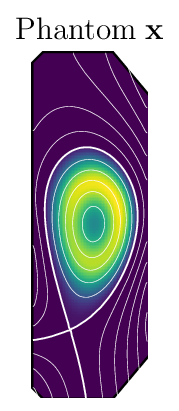}
         \caption{}
         \label{fig_phantom_gen_x}
     \end{subfigure}
\begin{subfigure}[b]{0.09\textwidth}
\centering
\raisebox{14pt}{\includegraphics[width=\textwidth]{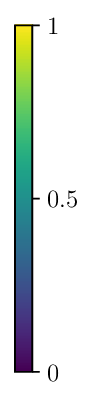}}
     \end{subfigure}
     \captionsetup{justification=raggedright,singlelinecheck=true}
     \caption{Phantom generation process: phantoms are obtained summing a randomly generated perturbation $\mathbf{x}_S$ to a radially decaying background $\mathbf{b}$.
     Figs. \ref{fig_phantom_gen0}-\ref{fig_phantom_gen3}: initial, intermediate and final anisotropic diffusion iterations performed to generate the perturbation $\mathbf{x}_S$; all iterations are normalized to $1$ for visualization purposes ($\widehat{\mathbf{x}}_i=\mathbf{x}_i/\max(\mathbf{x}_i)$). Fig. \ref{fig_phantom_gen_b}: radially decaying background $\mathbf{b}$. Fig. \ref{fig_phantom_gen_x}: phantom $\mathbf{x}$, with $\mathbf{x}=(\widehat{\mathbf{x}}_S+\mathbf{b})/\max(\widehat{\mathbf{x}}_S+\mathbf{b})$.} 
\label{fig_phantom_generation}
\end{figure}
Experimental observations from multiple tokamaks show that SXR emission can exhibit, due to a number of mechanisms, poloidal asymmetries, ring-like structures, peaked profiles, and other features, which can be described as perturbations upon a background SXR emission (see e.g. \cite{ingesson_soft_1998, vezinet_mesure_2013}). SXR tomography algorithms are typically validated on phantoms that feature profiles that are either Gaussian-like, hollow, peaked or exhibit high-field/low-field side (in/out as seen radially) asymmetries \cite{vezinet_mesure_2013, jardin_use_2017}. Inspired by these approaches, we construct a large set of $10^3$ SXR-like phantoms, each obtained by summing a calculated perturbation to a background profile. In Figure \ref{fig_phantom_generation}, we illustrate the phantom generation process, which we now describe in detail. We consider poloidally symmetric background emissions $\mathbf{b}$, defined as radially decaying flux surface functions whose peak coincides with the magnetic axis. The perturbations are generated by leveraging the diffusion equation analogy from Eq.\eqref{reaction-diffusion}, discussed in Section \ref{reaction_diffusion_section}. In particular, let $\psi$ be the poloidal magnetic flux function associated to a TCV magnetic equilibrium. Moreover, let $\mathbf{D}=\mathbf{D}(\psi, \alpha)$ be the anisotropic diffusion tensor introduced in Eq.\eqref{eq_diff_coeff}, defined as $\mathbf{D}_n=\mathbf{e}_{n,\parallel}\mathbf{e}_{n,\parallel}^{T}+\alpha\;\mathbf{e}_{n,\perp}\mathbf{e}_{n,\perp}^{T}$ for pixels $n=1,\ldots,N$;  the anisotropic parameter $\alpha<1$ determines the smoothing strength across flux surfaces, i.e., in the direction $\mathbf{e}_{n,\perp}$ oriented in the direction of the local gradient of $\psi$. Then, we obtain a SXR emissivity perturbation $\mathbf{x}_S$ by performing $S$ anisotropic diffusion iterations $\mathbf{x}_{n+1} = \mathbf{x}_n + \tau \boldsymbol{\nabla}\cdot(\mathbf{D}\boldsymbol{\nabla}\mathbf{x}_n)$, starting from a profile $\mathbf{x}_0$ corresponding to a point-wise source placed at a randomized location within the core.
\begin{figure}[t]
\begin{subfigure}[b]{0.8\textwidth}
\!\!\includegraphics[]{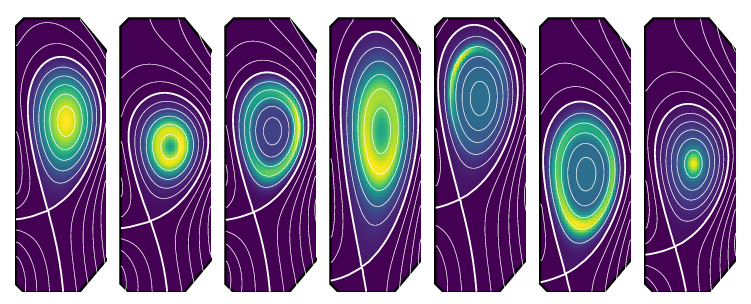}
\end{subfigure}
\hspace{1.3cm}
\begin{subfigure}[b]{0.09\textwidth}
\centering
\raisebox{-4pt}{\includegraphics[width=\textwidth]{phantom_generation_cbar.eps}}
     \end{subfigure}
\captionsetup{justification=raggedright,singlelinecheck=true}
\caption{Synthetic dataset: example of soft X-ray model phantoms.}
\label{fig_dataset}
\end{figure}
Finally, the phantom $\mathbf{x}$ is computed as the (normalized to 1) sum of the normalized perturbation $\widehat{\mathbf{x}}_S=\mathbf{x}_S/\max(\mathbf{x}_S)$ and background $\mathbf{b}$, i.e., $\mathbf{x}=(\widehat{\mathbf{x}}_S+\mathbf{b})/\max(\widehat{\mathbf{x}}_S+\mathbf{b})$. All computations are performed on a $240\times80$ pixel grid.\\
To increase the variability in the dataset, we introduce several additional sources of randomness. For each phantom, the magnetic equilibrium is obtained by rescaling and shifting, by random factors, a real TCV equilibrium $\psi_{_{TCV}}$. The peak amplitude of the background $\mathbf{b}$ is randomly selected, with bounds defined to satisfy $\max(\mathbf{b})\leq \max(\widehat{\mathbf{x}}_S)\leq 5\max(\mathbf{b})$. Both the number of anisotropic diffusion iterations and the anisotropic parameter $\alpha$ are randomly selected, with $10^3\leq S\leq 10^4$, $5\cdot 10^{-4}\leq\alpha\leq 5\cdot 10^{-1}$. Such bounds are picked as adequate, based on some trial runs, to obtain a sufficiently diverse and physically realistic dataset; perturbations are less localized for increasing values of $S$, smoother and less sharply anisotropic for larger values of $\alpha$. In Fig. \ref{fig_dataset}, we show a few of these generated phantoms. This large and diverse dataset can be used to test our tomographic pipeline over a wide range of emission profiles.
While not fully physically realistic, the class of phantoms defined by the described procedure includes all the phantom types typically used for SXR tomography validation, together with some of the most common experimentally observed emission patterns \cite{vezinet_mesure_2013, jardin_use_2017, ingesson_soft_1998}. This dataset is thus considered suitable for testing tomographic reconstruction algorithms.
\subsection{Phantom Analysis: Results}\label{results}We apply the Bayesian techniques presented in Section \ref{posterior} to the phantom dataset described in Section \ref{dataset}. We divide the $10^3$ phantoms in a set $\mathcal{S}_1$ of $100$ phantoms, used for hyperparameter tuning considerations (we will refer to Appendix \ref{appendix_hyperparam} for a more timely discussion), and a set $\mathcal{S}_2$ of $900$ phantoms, used to analyze the performance of the algorithms. For simplicity, we assume a line-integral approximation model \cite{ingesson_chapter_2008} for the geometry matrix $\mathbf{T}$. We model the noise $\boldsymbol{\varepsilon}\in\mathbb{R}^{M}$ corrupting the tomographic measurements in Eq.\eqref{ip_data} as Gaussian noise. In particular, let $\sigma_{N_1}$ and $\sigma_{N_2}$ be $5\%$ and $10\%$ of the average noiseless tomographic measurement value over the set $\mathcal{S}_1$, respectively; we consider three different noise models:
\vspace{-0.3cm}
\begin{itemize}
\item a moderate noise regime $N_1$ with $\varepsilon_m\sim\mathcal{N}(0, \sigma^2_{{N_1}})$, $m=1,\ldots,M$;
\vspace{-0.2cm}
\item a higher noise regime $N_2$ with $\varepsilon_m\sim\mathcal{N}(0, \sigma^2_{{N_2}})$, $m=1,\ldots,M$;
\vspace{-0.2cm}
\item a signal-dependent noise regime $N_3$ with $\varepsilon_m\sim\mathcal{N}(0, \sigma^2_{m}\,)$, with\\ $\sigma_m=\sigma_{N_1}+0.05\,(\mathbf{y}_{true})_m\,$, $m=1,\ldots,M$.
\end{itemize}
For reference, the noisy tomographic data showed in Fig.\ref{fig_diag} are obtained for noise model $N_2$.
At reconstruction time, only the noisy data $\mathbf{y}$ are provided to the algorithm; for each phantom, we attempt reconstruction from signals corrupted with each of the three noise models. For $N_1$, $N_2$, we assume that the channel-independent noise levels $\sigma_{N_1}$, $\sigma_{N_2}$, are known. For $N_3$, instead, we assume that only $\sigma_{N_1}$ is known, and the noise level for a given channel $m$ is estimated as $\hat{\sigma}_m=\sigma_{N_1}+0.05(\mathbf{y})_m\approx\sigma_m\,$; this allows to investigate robustness to noise mismodeling. For $N_1$, $\hat{\sigma}_m=\sigma_{N_1}$, and for $N_2$, $\hat{\sigma}_m=\sigma_{N_2}$, $\forall\,m$.
\\Throughout this Section, we consider the reconstruction model given by the posterior
\begin{equation}\label{eq_posterior_recon}
 \!\!p(\mathbf{x}\vert\mathbf{y})\propto \underbrace{\exp\!\!\,\Big(\!\!-\frac{1}{2}\big\Vert \mathbf{I}_{\widehat{\sigma}^{-\!1}}(\mathbf{y}-\mathbf{T}\mathbf{x})\big\Vert_2^2 \Big)}_{
 \displaystyle
 p(\mathbf{y}\vert\mathbf{x})}  \;\underbrace{\exp\!\!\,\Big(\!\!-\frac{\lambda}{2} \big\Vert \sqrt{\mathbf{D}(\psi,\alpha)}\boldsymbol{\nabla}\mathbf{x} \big\Vert_2^2 \Big)\,\mathcal{P}_{\mathbf{x}\geq0}}_{
 \displaystyle
 p(\mathbf{x})}
\end{equation}
where $\mathbf{I}_{\widehat{\sigma}^{-\!1}}$ is a diagonal matrix whose entries are given by $(\mathbf{I}_{\widehat{\sigma}^{-\!1}})_{mm}=\hat{\sigma}_{m}^{-1}$. The likelihood $p(\mathbf{y}\vert\mathbf{x})$ in Eq.\eqref{eq_posterior_recon} is Gaussian, as discussed in Remark \ref{remark_noise_model}; if the noise level does not depend on the signal itself, as in noise models $N_1$, $N_2$, the (negative log-) likelihood simplifies to Eq.\eqref{gaussian_loglikelihood}.
The prior $p(\mathbf{x})$ in Eq.\eqref{eq_posterior_recon} is a gradient penalizing, Gaussian smoothness prior (see Sections \ref{reaction_diffusion_section}, \ref{gaussian_prior_section}) combined with the positivity constraint prior $\mathcal{P}_{\mathbf{x}\geq0}$. We consider the same anisotropic diffusion prior class used in Section \ref{dataset}, and assume that $\psi$ is known; however, the anisotropic parameter $\alpha_{true}$ used to generate a given phantom is hidden to the reconstruction algorithm. We treat the anisotropic and regularization parameters $\alpha$ and $\lambda$ as tunable hyperparameters. We reconstruct only the core plasma region, leveraging the assumption, verified by our phantoms, that no emission emanates from outside it. All reconstructions are performed on a $120\times40$ pixel grid, corresponding to the standard grid currently employed for inversions at TCV \cite{sheikh_radcam-radiation_2022}. The reconstruction grid is coarser than the $240\times80$ phantom generation grid; since the tomographic measurements are computed by line-integration on the finer phantom grid, this choice introduces a source of model mismatch. This mitigates the risk of overestimating the performance of the algorithms due to the unrealistic assumption of perfect knowledge of the data-generating model, a practice sometimes referred to as an \textit{inverse crime} \cite{kaipio_statistical_2005}. To assess the reconstruction error, we then downsample the phantoms to the coarser grid.\\
Due to the positivity constraint in Eq.\eqref{eq_posterior_recon}, tomographic reconstruction and uncertainty quantification cannot be achieved using the closed-form expressions in Eq.\eqref{gaussian_mean_cov}, since $p(\mathbf{x}\vert\mathbf{y})$ is not Gaussian. As discussed in Remark \ref{remark_differentiable}, proximal techniques allow proper handling of the non-smooth positivity constraint term. In practice, this amounts to clipping negative values to zero at each iteration in the MAP computation in Eq.\eqref{gradient_descent}, and to introducing a smooth approximation of $\mathbbm{1}_{\mathbf{x}\geq0}$ for MCMC sampling in Eq.\eqref{ULA}; we refer to Appendix \ref{appendix_prox} for details. 
We remark once more that, while in this work we focus on the close-to-Gaussian model in Eq.\eqref{eq_posterior_recon}, the discussed techniques are general and can be applied to a wide class of imaging models (see Remark \ref{remark_differentiable} and Appendix \ref{appendix_prox}).\\
To perform MAP computation and MCMC sampling, given the noisy measurements $\mathbf{y}$ and the noise model $N_{i},\,i\in\{1,2,3\}$, we must select the prior hyperparameters $\lambda, \alpha$, together with the MCMC hyperparameters $K_b, K$, i.e., the number of burn-in and total MCMC iterations (see Section \ref{MCMC_section}). As mentioned at the beginning of this Section, we use the set $\mathcal{S}_1$ for considerations regarding their tuning. 
Hyperparameter tuning is important, as the performance of a reconstruction model can sensitively depend on it. As discussed in detail in Appendix \ref{appendix_hyperparam}, our analyses suggest that, for the phantom class and posterior model considered in this work, a sufficient tuning strategy consists in selecting $K_b=10^3$, $K=10^5$, $\lambda=10^{-1}$, and tuning $\alpha$ with a cross-validation-like approach based on the noisy tomographic data.\\
We now report the results of the phantom study. For each phantom in set $\mathcal{S}_2$, we generate a single realization of noisy tomographic data for each noise model. The obtained measurements $\mathbf{y}_{N_1}, \mathbf{y}_{N_2}, \mathbf{y}_{N_3}$, are provided as input for MAP computation and MCMC sampling, with hyperparameters $\lambda, \alpha, K_b, K$ selected as described above. We use the acquired ULA samples, obtained by Eq.\eqref{ULA}, to compute posterior-based Monte Carlo estimates of quantities of interest, using Eq.\eqref{MC_estimate}. In particular, we compute the quantities $\boldsymbol{\mu}_{_{ULA}},$ $\boldsymbol{\sigma}_{_{ULA}}$, $\mu^{P_{rad}}_{\,{\scriptscriptstyle ULA}}$, $\sigma^{{P_{rad}}}_{\,\scriptscriptstyle ULA}$, $\boldsymbol{\mu}^{\,peak}_{_{ULA}}$, and $\boldsymbol{\sigma}^{\,peak}_{_{ULA}}$, i.e., the pixel-wise mean and standard deviation, the mean and standard deviation of the total SXR radiated power, and the mean and standard deviation of the emissivity peak location, respectively. In the case of the quantities $\boldsymbol{\mu}^{\,peak}_{_{ULA}},\,\boldsymbol{\sigma}^{\,peak}_{_{ULA}}\,\in\mathbb{R}^2$, the statistics of the radial and vertical peak location are estimated independently.
To evaluate the accuracy of the Bayesian estimates, we compute for each phantom the following metrics, chosen as good general indicators of the quality of the estimates. For the MAP, we consider the RMSE $E_{_{MAP}}$, given by
\begin{equation}\label{rmse_def}
E_{_{MAP}}\,=\,\sqrt{\frac{1}{N} \sum_{i=0}^{N-1} \bigg( \big(\mathbf{x}\big)_i - \big(\mathbf{x}_{_{MAP}}\big)_i  \bigg)^2 }\quad,
\end{equation}
and the relative error on the SXR total radiated power $\delta_{_{MAP}}^{P_{rad}}=(P_{rad}^{\,\scriptscriptstyle MAP}-P_{rad}^{\,true})/P_{rad}^{\,true}$. For the MCMC estimates, we consider the RMSE $E_{{\boldsymbol{\mu}_{ULA}}}$, which is defined as in Eq.\eqref{rmse_def} but for $\boldsymbol{\mu}_{_{ULA}}$ instead of $\mathbf{x}_{_{MAP}}$, the relative error $\delta_{\boldsymbol{\mu}_{ULA}}^{P_{rad}}=(\mu^{P_{rad}}_{\,{\scriptscriptstyle ULA}}-P_{rad}^{\,true})/P_{rad}^{\,true}$, the distance $d_{peak}$ between the estimated and true peak location, and the fractions of core pixels (denoted by $f_{\sigma}$, $f_{2\sigma}$) such that the true emissivity value falls within one or two standard deviations $\boldsymbol{\sigma}_{_{ULA}}$ of the mean $\boldsymbol{\mu}_{_{ULA}}$.
\begin{figure}[]
\begin{subfigure}[]{\textwidth}
 \centering
 \includegraphics[width=\textwidth]{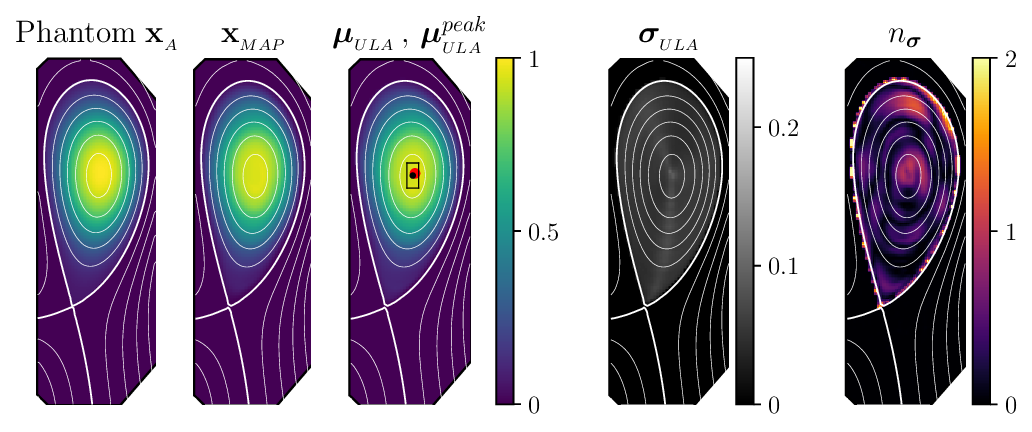}
     \end{subfigure}
     \vspace{-0.1cm}
     \\
     \begin{subfigure}[]{\textwidth}
 \centering
 \includegraphics[width=\textwidth]{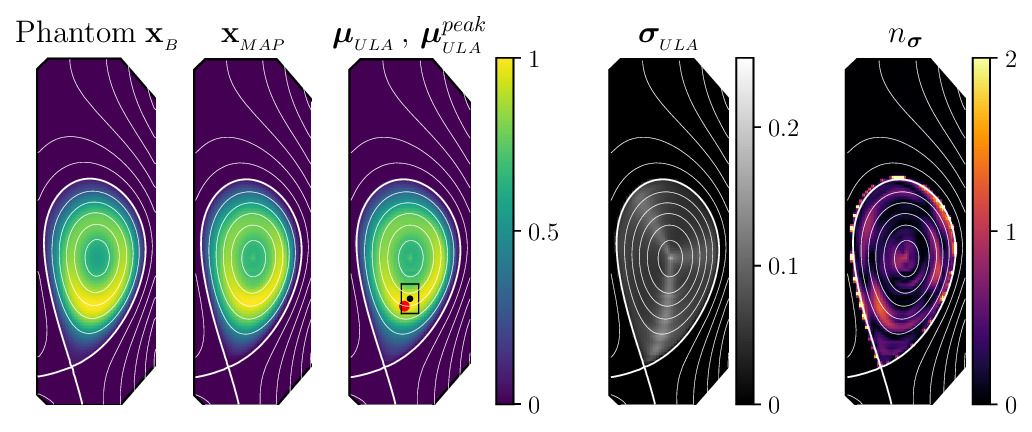}
     \end{subfigure}
     \vspace{-0.1cm}
     \\
     \begin{subfigure}[]{\textwidth}
 \centering
 \includegraphics[width=\textwidth]{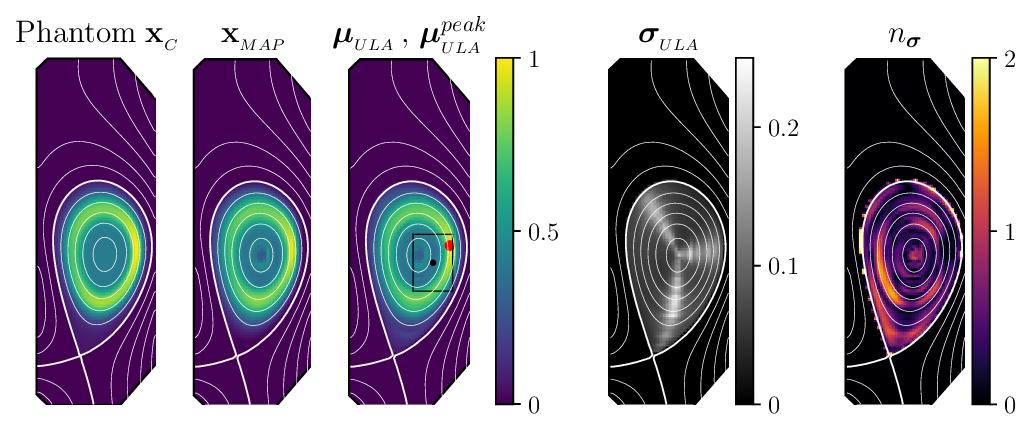}
     \end{subfigure}
     \vspace{-0cm}
     \\
$\phantom{space}$(a)$\qquad\quad\;\;\,$(b)$\qquad\quad\;\;\,$(c)$\qquad\qquad\qquad\quad\,$(d)$\qquad\qquad\qquad\;$(e)
     \\
\vspace{-0.4cm}
\caption{Bayesian analysis results for phantoms $\mathbf{x}_{_A}$, $\mathbf{x}_{_B}$, $\mathbf{x}_{_C}$, for noise model $N_2$. Figs.\ref{fig_results}a: phantoms. Figs.\ref{fig_results}b: MAPs.  Figs.\ref{fig_results}c: pixel-wise mean $\boldsymbol{\mu}_{_{ULA}}$; in black, $\boldsymbol{\mu}_{_{ULA}}^{peak}\pm\boldsymbol{\sigma}_{_{ULA}}^{peak}$, in red, true peak location. Figs.\ref{fig_results}d: pixel-wise variance $\boldsymbol{\sigma}_{_{ULA}}$.  Figs.\ref{fig_results}e: pixel-wise distance of the phantom from the mean, in terms of number of standard deviations, i.e.,  $n_{\boldsymbol{\sigma}}=\vert \mathbf{x}-\boldsymbol{\mu}_{_{ULA}}\vert / \boldsymbol{\sigma}_{_{ULA}}$.}
\label{fig_results}
\end{figure}
In Fig.\ref{fig_results} and Table \ref{table_results_sub}, we illustrate the results of the Bayesian analysis for three phantoms $\mathbf{x}_{_A}$, $\mathbf{x}_{_B}$, $\mathbf{x}_{_C}$, for tomographic data corrupted according to noise model $N_2$. In Figs.\ref{fig_results}b-\ref{fig_results}c, we can appreciate how both the MAP and mean estimators successfully reconstruct the main features of the corresponding phantoms; moreover, from Table \ref{table_results_sub}, we observe that $\mathbf{x}_{_{MAP}}$ and $\boldsymbol{\mu}_{_{ULA}}$ perform similarly well in terms of RMSE and $\delta^{P_{rad}}$. In Figs.\ref{fig_results}c, we additionally show the true emissivity peak location (red) and its estimated location $\boldsymbol{\mu}_{_{ULA}}^{peak}$ (black), together with the rectangular bounding box defined by $\boldsymbol{\mu}_{_{ULA}}^{peak}\pm\boldsymbol{\sigma}_{_{ULA}}^{peak}$, representing the uncertainty associated to the estimation. While the estimate $\boldsymbol{\mu}_{_{ULA}}^{peak}$ can be less accurate for some specific emissivity patterns, as for the shown phantom $\mathbf{x}_{{_C}}$, the predicted uncertainty $\boldsymbol{\sigma}_{_{ULA}}^{peak}$ is found to consistently provide a reliable statistical estimation of the peak location. In Figs.\ref{fig_results}d, we plot the pixel-wise standard deviation $\boldsymbol{\sigma}_{_{ULA}}$. Throughout our phantom analysis, we observe two interesting features: the magnitude of $\boldsymbol{\sigma}_{_{ULA}}$ depends on the estimated anisotropic parameter $\alpha$, while its spatial pattern exhibits limited variability across phantoms. Such features can be distinguished in Figs.\ref{fig_results}d. Indeed, the estimated uncertainty increases when moving from top to bottom of the figure, from the smoother phantom $\mathbf{x}_{_A}$ to the more anisotropic $\mathbf{x}_{_B}, \mathbf{x}_{_C}$. Furthermore, for all three phantoms, the standard deviation $\boldsymbol{\sigma}_{_{ULA}}$ is higher along two/three specific radial directions.
This is particularly apparent in the case of phantom $\mathbf{x}_{_C}$, due to the plotting scale, but all phantoms show some degree of this behavior. These two observations are ascribed to the uncertainty structure for our close-to-Gaussian reconstruction model in Eq.\eqref{eq_posterior_recon}. In the absence of the positivity constraint, $\boldsymbol{\sigma}_{_{ULA}}$ would simply be the square root of the diagonal of the posterior covariance matrix $\mathbf{\Sigma}_{post}=(\lambda\boldsymbol{\nabla}^T\mathbf{D}(\psi, \alpha)\boldsymbol{\nabla}+\mathbf{T}^T\mathbf{I}_{\widehat{\sigma}^{-\!2}}\mathbf{T})^{-1}$, as discussed in Section \ref{gaussian_prior_section}: the resulting $\boldsymbol{\sigma}_{_{ULA}}$ can then be shown to increase for decreasing values of $\alpha$, and its closed-form expression, for noise models $N_1$ and $N_2$, does not depend explicitly on the tomographic data $\mathbf{y}$, but only indirectly through the estimated value of $\alpha$. In the case of noise model $N_3$, $\boldsymbol{\sigma}_{_{ULA}}$ still conserves an explicit dependence on $\mathbf{y}$ through the term $\mathbf{I}_{\widehat{\sigma}^{-\!2}}$, but this dependence mainly affects its magnitude, without significantly affecting its spatial pattern.
\renewcommand{\arraystretch}{1.5}
\begin{table}[]
  \begin{tabularx}{0.99\textwidth}{c|Y|Y|Y|Y|c|c|c|}
    \cline{2-8}
     & $E_{_{MAP}}$ & $E_{{\boldsymbol{\mu}_{ULA}}}$ & $\delta_{_{MAP}}^{P_{rad}}$ & $\delta_{\boldsymbol{\mu}_{ULA_{}}}^{P_{rad}}$ & $d_{peak}$ & $f_{\sigma}$ & $f_{2\sigma}$ \\ \hline
    $\mathbf{x}_{_A}$ & $2.3\!\cdot\!10^{-2}$& $2.2\!\cdot\!10^{-2}$  & $-1.0\%$ & $-0.5\%$ & $1.3\,cm$ & $93.0\%$ & $98.4\%$\\ \hline
    $\mathbf{x}_{_B}$ &$3.3\!\cdot\!10^{-2}$ & $3.3\!\cdot\!10^{-2}$  & $+0.1\%$ & $+0.4\%$ & $4.0\,cm$& $91.6\%$ & $98.5\%$\\ \hline
    $\mathbf{x}_{_C}$ &$3.9\!\cdot\!10^{-2}$ & $4.6\!\cdot\!10^{-2}$  & $-0.6\%$ & $-0.3\%$ & $10.1\,cm$& $87.5\%$ & $98.3\%$ \\ \hline
\end{tabularx}
\captionsetup{justification=raggedright,singlelinecheck=true}
\caption{Metrics related to the Bayesian analysis results showed in Fig.\ref{fig_results}, for noise model $N_2$. We report: the RMSEs $E_{_{MAP}}$ and $E_{{\boldsymbol{\mu}_{ULA}}}$; the relative errors on the SXR radiated power $\delta_{_{MAP}}^{P_{rad}}$, $\delta_{\boldsymbol{\mu}_{ULA_{}}}^{P_{rad}}$; the distance $d_{peak}$ of $\boldsymbol{\mu}_{_{ULA}}^{peak}$ from the true peak; the fractions $f_{\sigma}$, $f_{2\sigma}$, of pixels such that $n_{\boldsymbol{\sigma}}\leq1$, $n_{\boldsymbol{\sigma}}\leq2$.}
\label{table_results_sub}
\end{table}
The presence of the positivity constraint does not significantly affect the two main findings: smaller values of $\alpha$ lead to larger $\boldsymbol{\sigma}_{_{ULA}}$,
and the spatial pattern of the quantified uncertainty exhibits limited dependence on the tomographic data, for reconstruction models similar to Eq.\eqref{eq_posterior_recon}. In particular, the pattern of $\boldsymbol{\sigma}_{_{ULA}}$ will be determined by the interplay between the prior term $\lambda\boldsymbol{\nabla}^T\mathbf{D}(\psi, \alpha)\boldsymbol{\nabla}$ and likelihood term $\mathbf{T}^T\mathbf{I}_{\widehat{\sigma}^{-\!2}}\mathbf{T}$.
Despite its mainly indirect dependence on the tomographic data $\mathbf{y}$, the estimated posterior standard deviation provides an effective way to quantify the uncertainty in the reconstruction. Indeed, $\boldsymbol{\sigma}_{_{ULA}}$ does not represent the local reconstruction error but, rather, a credibility measure of the local reconstructed value. In Figs.\ref{fig_results}e, we plot $n_{\boldsymbol{\sigma}}=\vert \mathbf{x} -\boldsymbol{\mu}_{_{ULA}}\vert / \boldsymbol{\sigma}_{_{ULA}}$, which measures the pixel-wise distance between the phantom and the posterior mean in terms of standard deviations. From Figs.\ref{fig_results}e and the values of $f_{\sigma}$, $f_{2\sigma}$ in Table \ref{table_results_sub}, we see that $\boldsymbol{\sigma}_{_{ULA}}$ can be used to define operative credible bounds $\boldsymbol{\mu}_{_{ULA}}\pm\boldsymbol{\sigma}_{_{ULA}}$ for the emissivity: for most pixels, the true emissivity value falls within one or two standard deviations of the mean.\\
\renewcommand{\arraystretch}{1.5}
\begin{table}[]
  \begin{tabularx}{0.99\textwidth}{c|Y|Y|Y|Y|c|c|c|}
    \cline{2-8}
    & $\widetilde{E}_{_{MAP}}$ & $\widetilde{E}_{{\boldsymbol{\mu}_{ULA}}}$ & $\widetilde{\delta}^{P_{rad}}_{_{MAP}}$ & $\widetilde{\delta}^{P_{rad}}_{{\;\boldsymbol{\mu}_{{ULA}}}}$ & $\widetilde{d}_{\,peak}$ & $\widetilde{f}_{\sigma}$ & $\widetilde{f}_{2\sigma}$ \\ \hline
    $N_1$ & $3.1\!\cdot\!10^{-2}$& $3.0\!\cdot\!10^{-2}$  & $-0.3\%$ & $+0.4\%$ & $4.3\,cm$ & $92.3\%$ & $98.5\%$\\ \hline
    $N_2$ &$4.1\!\cdot\!10^{-2}$ & $4.2\!\cdot\!10^{-2}$  & $-0.5\%$ & $+1.0\%$ & $5.6\,cm$& $88.8\%$ & $98.0\%$\\ \hline
    $N_3$ &$4.1\!\cdot\!10^{-2}$ & $4.3\!\cdot\!10^{-2}$  & $-1.0\%$ & $+0.1\%$ & $6.3\,cm$& $90.5\%$ & $98.0\%$ \\ \hline
\end{tabularx}
\captionsetup{justification=raggedright,singlelinecheck=false}
\caption{Phantom analysis summary: metrics related to Bayesian analysis results, for the studied noise models $N_1$, $N_2$, $N_3$. The reported quantities, the same included in Table \ref{table_results_sub}, are averages over the set of 900 phantoms $\mathcal{S}_2$.}
\label{table_results}
\end{table}
In Table \ref{table_results}, we report the average value of the metrics from Table \ref{table_results_sub} over the analyzed set of 900 phantoms $\mathcal{S}_2$, for the three considered noise regimes. For completeness, we report also the standard deviations of the same metrics over $S_2$, as Appendix \ref{appendix_results}. The results confirm the already discussed observations. For each noise model, the MAP and the posterior mean perform similarly well in terms of RMSE. As expected, the average RMSE is lower for the moderate noise regime $N_1$, compared to the higher noise regime $N_2$ and the signal-dependent noise regime $N_3$. Similarly, the average value of $d_{peak}$ is smaller for $N_1$, compared to the more challenging cases $N_2$, $N_3$. For all noise models, the fairly low average values of the RMSEs, $\delta^{P_{rad}}$ and $d_{peak}$ show that accurate estimates are typically achieved. The estimate of the total SXR radiated power, in particular, is very reliable; $P_{rad}^{\,true}$ is typically within one or two standard deviations $\sigma^{P_{rad}}_{\,{\scriptscriptstyle ULA}}$ of the mean $\mu^{P_{rad}}_{\,{\scriptscriptstyle ULA}}$. Moreover, crucially, the average values of $f_{\sigma}$ and $f_{2\sigma}$ confirm that posterior-based uncertainty quantification allows to provide meaningful credible bounds associated to the reconstruction, with approximately $90\%$ and $98\%$ of the core pixels falling within one and two standard deviations of the posterior mean, respectively.
In terms of computation time, when performed on a single core of the workstation specified at the beginning of Section \ref{application}, MAP computation and MCMC sampling took approximately $0.5s$ and $50s$ per phantom, respectively.
MAP computation provides a faster answer, which can be enriched by more advanced Bayesian analyses via MCMC sampling. The results demonstrate that, by adopting the rich Bayesian approach, we can provide principled statistical estimates of quantities of interest together with their associated uncertainty. This is obtained by statistical inference based on a clear, interpretable model of the acquisition system, the noise corrupting the data and our prior belief on the emissivity properties. The approach is demonstrated to be robust with respect to moderate model mismatch and, as showed from noise model $N_3$, noise mismodeling. The analysis on the large and diverse corpus of physically-motivated phantoms $\mathcal{S}_2$ confirms that meaningful inference can be made using the well-understood, simple and robust algorithms introduced in Sections \ref{MAP_section}, \ref{MCMC_section}.
\subsection{Sparse-view tomography: limitations}\label{limitations}In Section \ref{results}, we showed that Bayesian inference is a powerful tool. However, in the sparse-view, limited-angle regime characterizing plasma tomography, there are limits to what any technique can achieve. It is important to identify and highlight them, and remain aware of some of the pitfalls that can be encountered.
Due to the sparseness and noisiness of the tomographic data, potential noise mismodeling, and other sources of model mismatch, perfect reconstruction cannot be achieved.
It is known that smoothing priors, such as those used in this work, tend to over-smooth emissivity profiles \cite{ingesson_chapter_2008}. Such effects can be more or less pronounced depending on the actual emissivity distribution. To discuss these limitations, we consider an artificial diagnostic $\mathcal{T}_{art}$, composed of $10^4$ randomly chosen lines of sight, that are allowed to disregard the technical constraints dictated by the locations of the physical ports on TCV. Line of sight configurations, apart from being plotted as in Fig.\ref{fig_diag}, can also be visualized in projection space. This is done \cite{kak_principles_2001} by parametrizing each line of sight in terms of its distance $p$ from a chosen fixed point, and the angle $\theta$ describing its steepness.
\begin{figure}[t!]
\begin{subfigure}[]{0.5\textwidth}
 \centering
 \includegraphics[width=\textwidth]{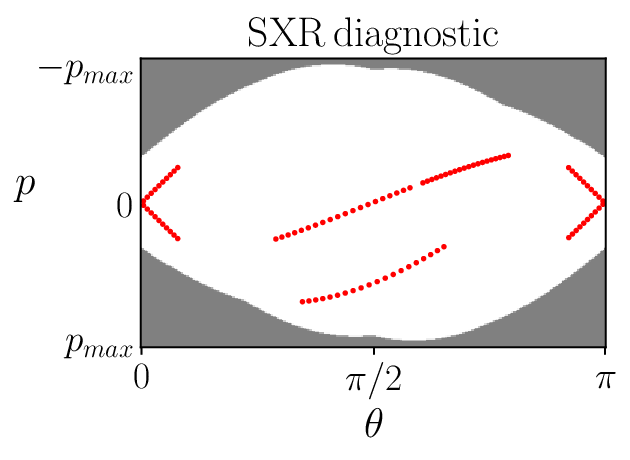}
     \end{subfigure}
     \hspace{0.1cm}
\begin{subfigure}[]{0.49\textwidth}
 \centering
 \includegraphics[width=\textwidth]{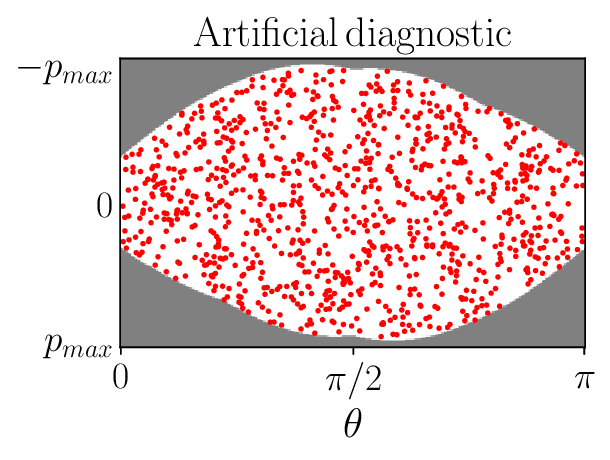}
     \end{subfigure}
\vspace{-0.5cm}
\captionsetup{justification=raggedright,singlelinecheck=false}
\caption{Lines of sight configurations in projection space. Red dots, white regions, and shaded areas, represent lines of sight, admissible $(p,\theta)$ values, and $(p,\theta)$ values giving lines which fall outside the TCV vessel, respectively. To the left, TCV's SXR system. To the right, the artificial diagnostic $\mathcal{T}_{art}$ composed of $10^4$ lines of sight, randomly sampled from the admissible $(p,\theta)$ domain; only $10\%$ of the total lines are plotted, for visualization purposes.}
\label{fig_limitations_sinogram}
\end{figure}
\begin{figure}[t!]
\begin{subfigure}[]{\textwidth}
 \centering
 \includegraphics[width=\textwidth]{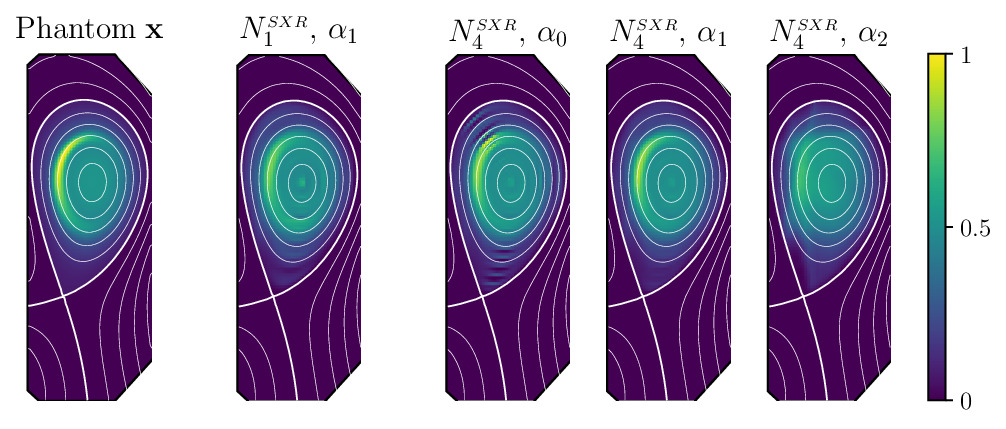}
     \end{subfigure}
     \vspace{-0.1cm}
     \\
\begin{subfigure}[]{\textwidth}
 \centering
 \includegraphics[width=\textwidth]{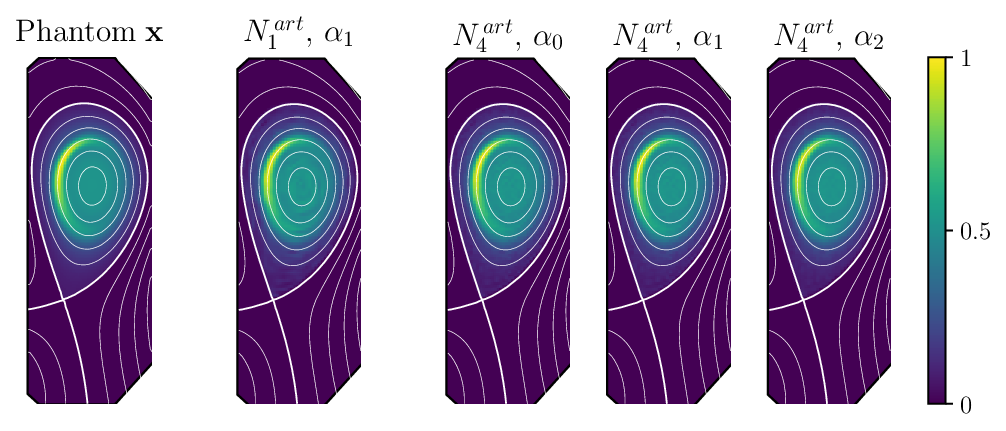}
     \end{subfigure}
     \\
     \vspace{0.2cm}
     \\
$\phantom{spaci}$(a)$\qquad\qquad\quad\;\;$(b)$\qquad\qquad\quad\;\;$(c)$\qquad\qquad$(d)$\qquad\qquad\;$(e)
     \\
     \vspace{-0.3cm}
\captionsetup{justification=raggedright,singlelinecheck=false}
\caption{Sparse-view tomography limitations: comparison between TCV's SXR diagnostic (top row) and the artificial diagnostic $\mathcal{T}_{art}$ (bottom row). In Figs.\ref{fig_limitations}a, the considered phantom. In Figs.\ref{fig_limitations}b, the MAPs obtained for noise model $N_1$, with anisotropic parameter $\alpha_1=10^{-2}$. In Figs.\ref{fig_limitations}c-\ref{fig_limitations}e, the MAPs obtained for the small noise regime $N_4$, with $\alpha_0=\!10^{-4}$, $\alpha_1=\!10^{-2}$, $\alpha_2=\!10^{0}$.}
\label{fig_limitations}
\end{figure}
 \renewcommand{\arraystretch}{1.5}
 \begin{table}[]
  \begin{tabularx}{0.975\textwidth}{c|Y|Y|Y|Y|Y|}
    \cline{2-6}
    & Diagnostic & $N_1, \;\alpha_1$ & $N_4,\; \alpha_0$ & $N_4, \;\alpha_1$ & $N_4, \;\alpha_2$ \\ \hline
    $E_{_{MAP}}$ & SXR & $4.6\cdot10^{-2}$& $6.9\cdot10^{-2}$  & $2.9\cdot10^{-2}$ & $6.8\cdot10^{-2}$ \\ \hline
    $\delta_{_{MAP}}^{P_{rad}}$ & SXR & $+0.9\%$& $+0.7\%$  & $+0.5\%$ & $+0.1\%$ \\ \hline
    $E_{_{MAP}}$ & $\mathcal{T}_{art}$ & $2.1\cdot10^{-2}$& $1.6\cdot10^{-2}$  & $1.5\cdot10^{-2}$ & $1.0\cdot10^{-2}$ \\ \hline
    $\delta_{_{MAP}}^{P_{rad}}$ & $\mathcal{T}_{art}$ & $+0.2\%$& $+0.3\%$  & $+0.3\%$ & $+0.3\%$\\ \hline
\end{tabularx}
\captionsetup{justification=raggedright,singlelinecheck=false}
\caption{Comparison between TCV's SXR diagnostic and the artificial diagnostic $\mathcal{T}_{art}$. RMSEs $E_{_{MAP}}$ and relative errors on SXR radiated power $\delta_{_{MAP}}^{P_{rad}}$, for the same phantom, noise models, and $\alpha$ values considered in Fig.\ref{fig_limitations}.}
\label{table_limitations}
\end{table}
$\!\!\!\!\!\!\!\!$In Fig.\ref{fig_limitations_sinogram}, we show the projection space representation of the SXR diode system installed at TCV, and of the artificial diagnostic $\mathcal{T}_{art}$. We compute the distances $p$ with respect to the vessel cross-section geometric center, with $\theta$ the angle between horizontal and the normal to the line of sight \cite{kak_principles_2001}. Following the convention also adopted in \cite{ingesson_chapter_2008}, $(p,\theta)\in\mathbb{R}\!\times\![0,\pi]$; negative values of $p$ correspond to lines of sight lying below the chosen fixed point. Clearly, $\mathcal{T}_{art}$ covers the projection space better than the SXR diagnostic, resulting in improved tomographic capabilities. In particular, since our $120\times40$ reconstruction grid is composed of $4800$ pixels, the tomographic reconstruction problem for data acquired with $\mathcal{T}_{art}$ is now overconstrained, compared to the severely underconstrained regime characterizing the TCV's SXR diagnostic. To discuss some fundamental Bayesian concepts, we compare the performance of these configurations in Fig.\ref{fig_limitations} and Table \ref{table_limitations}. For simplicity, we only focus on MAP reconstruction. In Figs.\ref{fig_limitations}b, we show that, for the moderate noise regime $N_1$, the reconstruction of the SXR diagnostic, with $\alpha_1=\alpha_{CV}=10^{-2}$, smooths out the fine poloidally asymmetric feature in the considered phantom $\mathbf{x}$. Similar behavior is observed for several other phantoms. Importantly, we remark that, while the MAP might be less accurate for some of the phantoms, the Bayesian approach still typically yields good probabilistic coverage of the local emissivity value; for the considered phantom, indeed, it holds $f_{\sigma}=87.3\%$, $f_{2\sigma}=96.9\%$. As expected, from Figs.\ref{fig_limitations}b and Table \ref{table_limitations} we see that the improved coverage of $\mathcal{T}_{art}$ results in a significantly improved reconstruction. However, even in the moderate noise regime $N_1$, the prior plays an important role for $\mathcal{T}_{art}$, similarly to what happens for the more limited SXR diagnostic: ill-estimation of $\alpha$ considerably deteriorates the reconstruction quality (not shown).
Let us now introduce the small noise regime $N_4$, with signal-independent noise of intensity $\sigma_{N_4}=0.2\,\sigma_{N_1}$, i.e., $1\%$ of the average tomographic measurements (see Section \ref{results}). This is unrealistically low for a real experimental system. In Figs.\ref{fig_limitations}c-\ref{fig_limitations}e, we show the obtained MAPs, for noise model $N_4$, for three values of $\alpha$. For the underdetermined case of the SXR diagnostic, even in the small noise limit, the chosen prior strongly affects the reconstruction; while the reconstruction is rather good for $\alpha=10^{-2}$, artifacts arise for the lower value $\alpha=10^{-4}$, and the emissivity is massively oversmoothed for the higher value $\alpha=10^0$. For the overdetermined case of $\mathcal{T}_{art}$, all values of $\alpha$ perform similarly well, and achieve excellent results. The RMSEs corresponding to each MAP are reported in Table \ref{table_limitations}. Interestingly, the estimation of the total SXR radiated power is far less sensitive to hyperparameter ill-estimation than the RMSEs, as confirmed by the relative errors reported in Table \ref{table_limitations}.\\
These observations, related to the posterior measure behavior in the limit of small noise, are well-known features of the Bayesian paradigm \cite{stuart_inverse_2010}. In the underdetermined setting, the prior significantly influences the solution even for vanishing noise. In the overdetermined case, instead, the prior does not play any role for vanishing noise; posterior-based uncertainty converges to zero, and the entire probability mass concentrates around the solution of the least-squares problem in Eq.\eqref{least_squares}. Unfortunately, plasma tomography problems are typically severely underdetermined.
As a consequence, careful modeling of the prior is necessary.
The sparse-view nature of plasma tomography implies that inference results will be sensitive to the prior and its hyperparameters; the noisiness of the data makes the situation worse and further limits tomographic capabilities. This explains why it remains crucial to provide a statistical answer to the challenging problem of plasma tomographic reconstruction. A single reconstruction, with no attempt at uncertainty quantification, cannot provide an accurate and complete answer. Here lies the attractiveness of the Bayesian approach, due to its inherent ability to quantify the uncertainty associated with the estimation of quantities of interest.

\section{Summary, Conclusion and Next Steps}\label{conclusion}In this work, we discussed a unifying framework for Bayesian inference from plasma imaging tomographic data. We showed that many popular plasma emission tomographic techniques (such as isotropic/anisotropic approaches based on generalized Tikhonov or MFI regularization, Gaussian Process Tomography, and Maximum Likelihood tomography) can be naturally understood within a common Bayesian framework. Operating within such a framework, we leveraged well-studied optimization and inference concepts to propose and describe in detail a versatile pipeline that relies on simple, robust, and efficient algorithms. In particular, we proposed the unadjusted Langevin algorithm for efficient posterior sampling. We applied this pipeline to the problem of SXR imaging at TCV, demonstrating that the proposed approach enables principled, useful inference from noisy, sparse-view tomographic data. We validated the pipeline by conducting a phantom-based study over a large set of physically realistic SXR model phantoms. We showed that the pipeline leads to a statistically reliable estimation of the main features of the phantoms (in terms of shape, asymmetry, anisotropy, total SXR radiated power and peak location) and demonstrated the robustness of the approach with respect to moderate noise mismodeling and model mismatch. We showed that MAP computation can be complemented by MCMC methods for posterior sampling, which allow uncertainty quantification and complex statistical analysis but are more computationally expensive; such a complementary vision, with frameworks combining quick MAP-based analyses with detailed MCMC-based computations on data of particular interest, is becoming increasingly popular. 
Finally, we highlighted some of the inherent limitations of sparse-view tomography and the crucial role played by the prior in such an underdetermined setting: careful prior modeling is clearly essential. Moreover, we stressed the importance of adopting a probabilistic approach allowing to quantify the uncertainty associated to any estimate, which is made all the more vital by the sparse-view nature of plasma tomography. Open access is provided to the computational routines implemented in the context of this work, in an effort towards reproducibility and reusability: the link to the project repository can be found at the beginning of Section \ref{application}. Instructions are provided to generate the data used in this work, which are made available for further tomography studies. This software and data availability policy is chosen to facilitate building upon the proposed framework.\\
The unifying view on plasma tomography fostered by the Bayesian approach is interesting both from a conceptual and a practical standpoint. It promotes a general, statistically principled, and consistent approach to tomographic reconstruction and uncertainty quantification, not restricted to a single class of likelihood and prior models associated with a specific tomographic technique. Moreover, importantly, framing the plasma emission tomographic problem in the context of modern convex optimization and Bayesian analysis enables cross-fertilization from the broader computational imaging community, which has been making significant advances in recent years. We strongly believe that there is much to benefit from investigating the application of some of these developments to plasma imaging.\\
For future work, we identify an applied and a methodological axis. From the application viewpoint, we will assess the scientific potential of the proposed pipeline by its application to tomographic reconstruction from the widest possible range of experimental TCV data. In particular, we will broaden the scope from SXR imaging to include bolometry and, potentially, visible-light imaging applications. This can be readily achieved within the generality of the proposed inference framework. Future studies could also investigate principled comparisons between techniques that have been typically viewed as distinct and unrelated, e.g., between traditional smoothness-promoting techniques and a fully Bayesian version of the ML method, as mentioned in Remark \ref{remark_noise_model}: this is now possible within the proposed unifying framework. From the methodological and modeling points of view, future studies could investigate priors that are different from the classical smoothness approaches traditionally considered for plasma imaging. Specifically, it would be very interesting to consider sparsity-promoting priors \cite{bertero_introduction_2020, cai_uncertainty_2018-1} and recently popularized deep learning-based priors \cite{laumont_bayesian_2022}.

\section*{Acknowledgements}
This work was supported in part by the Swiss National Science Foundation. This work has been carried out within the framework of the EUROfusion Consortium, via the Euratom Research and Training Programme (Grant Agreement No 101052200 — EUROfusion) and funded by the Swiss State Secretariat for Education, Research and Innovation (SERI). Views and opinions expressed are however those of the author(s) only and do not necessarily reflect those of the European Union, the European Commission, or SERI. Neither the European Union nor the European Commission nor SERI can be held responsible for them.

\bibliographystyle{unsrt}
\typeout{}
\bibliography{tomo_fusion}

\appendix
\section*{Appendix}

\renewcommand{\thesubsection}{\Alph{subsection}}

\subsection{Phantom Analysis: supplementary results}
\label{appendix_results}
In this appendix, we report in Table \ref{table_results_std} the standard deviations, computed over the phantom set $\mathcal{S}_2$, of the metrics employed in Section \ref{results} to analyze the performance of the applied inference algorithms. We refer to Section \ref{results} for a definition of the reported quantities. Table \ref{table_results_std} completes the information provided in Table \ref{table_results}, which reported the average value of the metrics over the same phantom set.
\renewcommand{\arraystretch}{1.6}
\begin{table}[h!]
  \begin{tabularx}{0.99\textwidth}{c|Y|Y|Y|Y|c|c|c|}
    \cline{2-8}
    & $s(E_{_{MAP}})$ & $s\big(E_{{\boldsymbol{\mu}_{ULA}}}\big)$ & $s\big(\delta^{P_{rad}}_{_{MAP}}\big)$ & $s\big(\delta^{P_{rad}}_{{\;\boldsymbol{\mu}_{{ULA}}}}\big)$ & $s\big(d_{\,peak}\big)$ & $s\big(f_{\sigma}\big)$ & $s\big(f_{2\sigma}\big)$ \\ \hline
    $N_1$ & $0.9\!\cdot\!10^{-2}$& $0.9\!\cdot\!10^{-2}$  & $0.6\%$ & $1.0\%$ & $3.7\,cm$ & $4.4\%$ & $1.1\%$\\ \hline
    $N_2$ &$1.3\!\cdot\!10^{-2}$ & $1.3\!\cdot\!10^{-2}$  & $1.1\%$ & $1.9\%$ & $4.3\,cm$& $5.9\%$ & $1.6\%$\\ \hline
    $N_3$ &$1.2\!\cdot\!10^{-2}$ & $1.3\!\cdot\!10^{-2}$  & $0.9\%$ & $1.4\%$ & $4.6\,cm$& $5.1\%$ & $1.7\%$ \\ \hline
\end{tabularx}
\captionsetup{justification=raggedright,singlelinecheck=false}
\caption{Phantom analysis summary: standard deviation $s$, over the set of 900 phantoms $\mathcal{S}_2$, of the metrics related to Bayesian analysis results, for the three studied noise models $N_1$, $N_2$, $N_3$.}
\label{table_results_std}
\end{table}

\subsection{Handling Non-Smooth Terms: Proximal Operators and Moreau-Yosida Approximations}
\label{appendix_prox}
As mentioned in Remark \ref{remark_differentiable} in Section \ref{posterior}, the discussed Bayesian inference techniques for MAP computation and MCMC sampling can be adapted to models featuring non-continuously differentiable log-posteriors. In this appendix, we briefly introduce the proximal techniques allowing this, referring the interested reader, with references, to more detailed explanations. Many interesting imaging models feature \cite{combettes_proximal_2011} convex negative log-posteriors composed of a smooth and a non-smooth term, 
\begin{equation}\label{f_g_log_posterior}
    -\log p(\mathbf{x}\vert\mathbf{y}) = f(\mathbf{x}) + g(\mathbf{x})\;,
\end{equation}
where $f:\mathbb{R}^N\to\mathbb{R}$ is a convex, smooth function with $\beta$-Lipschitz continuous gradient (see Eq.\eqref{lipschitz_continuity}), and $g:\mathbb{R}^N\to\mathbb{R}$ is a convex, non-smooth function (satisfying the additional technical conditions of being proper and lower semi-continuous, or, equivalently, closed \cite{parikh_proximal_2014}). For functions $g$ with such properties, the \emph{proximal operator} $\mathrm{prox}_{\nu g}$ of the scaled function $\nu g$, with $\nu>0$, is defined as
\begin{equation}\label{prox_op}
   \mathrm{prox}_{\nu g}(\mathbf{x}) = \argmin_{\mathbf{z}\in\mathbb{R}^N}\;\Big(g(\mathbf{z}) + \frac{1}{2\nu}\Vert\mathbf{x}-\mathbf{z}\Vert_2^2 \Big)\;;
\end{equation}
the proximal operator $\mathrm{prox}_{\nu g}:\mathbb{R}^N\to\mathbb{R}^N$ thus maps $\mathbf{x}$ to the unique solution of the minimization problem in Eq.\eqref{prox_op} \cite{combettes_proximal_2011, parikh_proximal_2014}. The proximal operator in Eq.\eqref{prox_op} can often be evaluated efficiently \cite{combettes_proximal_2011, parikh_proximal_2014}. The proximal mapping can be seen as a generalization of the concept of projection \cite{combettes_proximal_2011, parikh_proximal_2014}. Indeed, if $g=\mathbbm{1}_{\mathcal{C}}$ is the indicator function of a convex, closed set $\mathcal{C}$, i.e.,
\begin{equation}\label{indicator_fct}
    \mathbbm{1}_{\mathcal{C}}(\mathbf{x})=\begin{cases}
			0 & \text{if $\mathbf{x}\in\mathcal{C}$}\\
            +\infty & \text{if $\mathbf{x}\notin\mathcal{C}$}
		 \end{cases}
\end{equation}
then $\mathrm{prox}_{\nu g}$ is simply the orthogonal projection operator on $\mathcal{C}$.\\
The proximal operator is intimately related to the $\nu$-Moreau-Yosida envelope \cite{parikh_proximal_2014} of the function $g$, which is defined as
\begin{equation}\label{my_envelope}
    g^{\nu}(\mathbf{x}) = \inf_{\mathbf{z}\in\mathbb{R}^N}\;\Big(g(\mathbf{z}) + \frac{1}{2\nu}\Vert\mathbf{x}-\mathbf{z}\Vert_2^2 \Big)\;.
\end{equation}
The Moreau-Yosida envelope is a smooth approximation of the non-smooth $g$; it can be made arbitrarily close to $g$ by reducing $\nu$. The function $g^{\nu}$ is continuously differentiable with $\nu^{-1}$-Lispchitz continuous gradient. Moreover, it can be shown that
\begin{equation}\label{grad_of_my_envelope}
   \mathrm{prox}_{\nu g}(\mathbf{x}) = \mathbf{x} - \nu \nabla g^\nu(\mathbf{x})\;.
\end{equation}
Proximal algorithms have become a cornerstone of computational imaging and signal processing at large. In particular, minimization of the functional $f+g$ in Eq.\eqref{f_g_log_posterior}, corresponding to MAP computation in the Bayesian setting, can be achieved by the \emph{proximal gradient descent} algorithm \cite{combettes_proximal_2011, parikh_proximal_2014}
\begin{equation}\label{prox_grad_desc}
    \mathbf{x}_{k+1} =  \mathrm{prox}_{\nu g}\Big(\mathbf{x}_k - \nu\nabla f(\mathbf{x}_k)\Big),\quad \mathbf{x}_0=\mathbf{x}_0,\quad k\geq0\;,
\end{equation}
i.e., by alternating gradient steps on the smooth $f$, just as in Eq.\eqref{gradient_descent} for standard gradient descent, with proximal steps on the non-smooth $g$. Convergence is guaranteed
if $\nu\leq1/\beta$ \cite{combettes_proximal_2011, parikh_proximal_2014}, just as for Eq.\eqref{gradient_descent}. Furthermore, from Eq.\eqref{grad_of_my_envelope}, we see that the proximal step in Eq.\eqref{prox_grad_desc} can be interpreted as a gradient descent step performed on the Moreau-Yosida envelope $g^{\nu}$.\\
Proximal MCMC methods, allowing sampling from the posterior distribution, have also been recently proposed: in this work, we considered the \emph{Moreau-Yosida regularized Unadjusted Langevin Algorithm} (MYULA) \cite{durmus_efficient_2018}
\begin{equation}
    \label{eq_myula}
    \mathbf{x}_{k+1} = \mathbf{x}_k + \tau \Big( \nabla f(\mathbf{x}_k) + \nabla g^{\nu}(\mathbf{x}_k)\Big) + \sqrt{2\tau}\,\mathbf{z}_k,\quad k\geq0.
\end{equation}
Compared to ULA in Eq.\eqref{ULA}, the proximal MCMC method in Eq.\eqref{eq_myula} replaces the negative log-posterior $f+g$ by the smooth approximation $f+g^{\nu}$. Since $\nabla f$ is $\beta$-Lipschitz and $\nabla g^{\nu}$ is $\nu^{-1}$-Lipschitz, $\nabla (f+g^{\nu})$ is $L$-Lipschitz with $L\leq \beta + \nu^{-1}$ \cite{durmus_efficient_2018}: as a consequence, similarly to that discussed in Section \ref{MCMC_section} for ULA, the condition $\tau\leq 1/L$ is required for stability.\\
In this work, we employ the reconstruction model in Eq.\eqref{eq_posterior_recon}. While the exponential terms in Eq.\eqref{eq_posterior_recon} are smooth, the positivity constraint prior $\mathcal{P}_{\mathbf{x}\geq0}=\exp(-\mathbbm{1}_{\mathbf{x}\geq0})$ leads to a non-smooth, proximable term in the negative log-posterior. Therefore, we split $-\log p(\mathbf{x}\vert\mathbf{y})$ into a smooth and non-smooth component as in Eq.\eqref{f_g_log_posterior}, with $g=\mathbbm{1}_{\mathbf{x}\geq0}$ defined as in Eq.\eqref{indicator_fct} for $\mathcal{C}=\{\mathbf{x}\in\mathbb{R}^N: x_i\geq0,\,\forall i=1,\ldots,N\}$. Then, we compute the MAP estimate using the proximal gradient descent algorithm in Eq.\eqref{prox_grad_desc}, and sample from the posterior using the MYULA algorithm in Eq.\eqref{eq_myula}. The proximal operator of the positivity constraint, when applied to a vector $\mathbf{x}\in\mathbb{R}^N$, simply clips to zero all the negative entries of $\mathbf{x}$. This operation corresponds to a projection onto the positive orthant, which is a convex, closed set.\\
The algorithms discussed in this appendix allow advanced Bayesian statistical analysis for non-smooth log-posteriors that arise in many imaging models. In this work, we employed these algorithms to enforce a positivity constraint on the reconstruction, as is customary in plasma tomography. These techniques allow the treatment of such constraints in a principled way both for MAP computation and MCMC sampling. The scope of the discussed techniques is, however, much broader. In particular, the algorithms introduced in this appendix can be applied to models featuring sparsity-promoting priors and for a popular class of recently proposed deep learning based priors. We refer to \cite{durmus_efficient_2018, cai_uncertainty_2018-1} and \cite{laumont_bayesian_2022} for discussions of Bayesian computations in the case of sparsity-promoting and deep learning based priors, respectively. These represent interesting and exciting research avenues, as yet unexplored by the plasma tomography community, that are planned for investigation in future work.

\subsection{Hyperparameter Tuning Strategy}
\label{appendix_hyperparam}
\begin{figure}[t]
\begin{subfigure}[b]{0.315\textwidth}
\!\!\includegraphics[width=\textwidth]{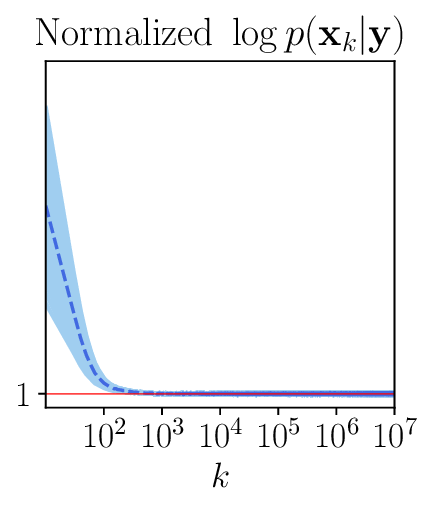}
\vspace{-0.26cm}
\caption{}
\label{fig_ula_tuning_logp}
\end{subfigure}
\hspace{-0.4cm}
\begin{subfigure}[b]{0.34\textwidth}
\!\!\includegraphics[width=\textwidth]{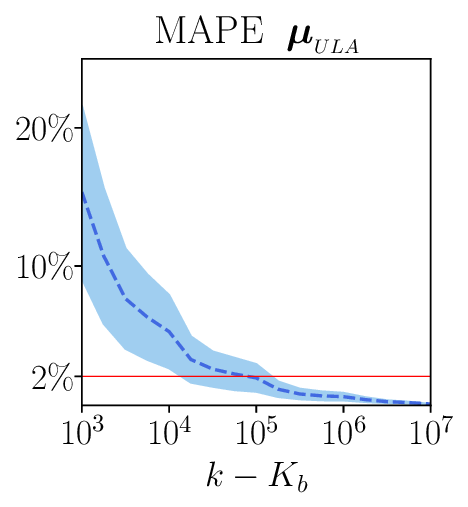}
\vspace{-0.22cm}
\caption{}
\label{fig_ula_tuning_mape_mean}
\end{subfigure}
\hspace{-0.3cm}
\begin{subfigure}[b]{0.34\textwidth}
\!\!\includegraphics[width=\textwidth]{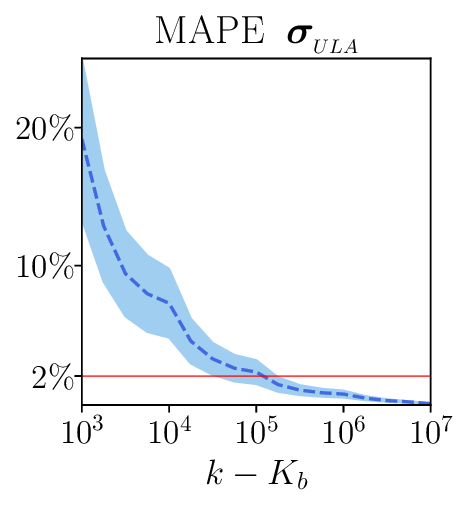}
\vspace{-0.2cm}
\caption{}
\label{fig_ula_tuning_mape_std}
\end{subfigure}
\vspace{-0.25cm}
\captionsetup{justification=raggedright,singlelinecheck=true}
\caption{ULA hyperparameter tuning. Dashed lines and shaded areas represent the average and standard deviation over the phantom set $\mathcal{S}_1$.\\Fig. \ref{fig_ula_tuning_logp}: log-posterior values normalized by their plateau value, showing that $K_b\approx10^3$ is a sufficient number of burn-in iterations. Figs. \ref{fig_ula_tuning_mape_mean}, \ref{fig_ula_tuning_mape_std}: MAPEs committed approximating $\boldsymbol{\mu}_{_{ULA}}^{10^7}$ and $\boldsymbol{\sigma}_{_{ULA}}^{10^7}$ by $\boldsymbol{\mu}_{_{ULA}}^{k-K_b}$ and $\boldsymbol{\sigma}_{_{ULA}}^{\,k-K_b}$.
}
\label{fig_ula_tuning}
\end{figure}

In this appendix, we provide details upon the tuning strategy adopted for the prior hyperparameters $\lambda, \alpha$, and MCMC hyperparameters $K_b, K$. As discussed in Section \ref{results}, we study hyperparameter tuning on a dedicated set of 100 phantoms $\mathcal{S}_1$. We then rely upon the selected tuning strategies in Section \ref{results}, when testing the proposed pipeline on the remaining set of $900$ model phantoms $\mathcal{S}_2$: we exclude the phantoms in $\mathcal{S}_1$ from the analysis to avoid leaking hyperparameter tuning ``training'' information into the results' analysis.\\
First, let us consider the regularization parameter $\lambda$ and the anisotropic parameter $\alpha$. Regularization parameter tuning is a rich and well-studied topic (see e.g. \cite{fernandez_vidal_maximum_2020} and references therein); we refer to \cite{ingesson_chapter_2008} for a discussion within the context of plasma tomography. Moreover, in the particular case of Gaussian posteriors, hyperparameters are typically tuned by marginal likelihood maximization \cite{rasmussen_gaussian_2005, moser_gaussian_2022}.
\begin{figure}[t]
\hspace{0.5cm}
\begin{subfigure}[b]{0.45\textwidth}
\!\!\includegraphics[width=\textwidth]{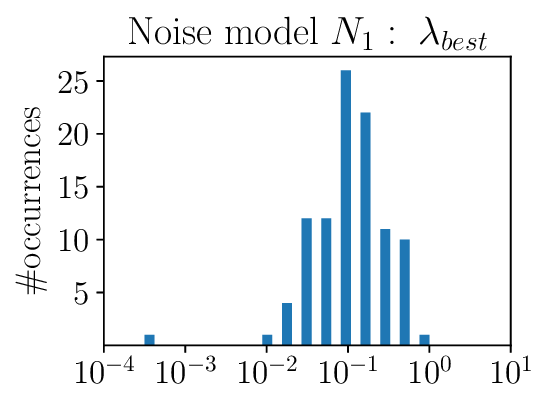}
\vspace{-0.3cm}
\caption{}
\label{fig_reg_param_tuning}
\end{subfigure}
\hspace{0.5cm}
\begin{subfigure}[b]{0.4\textwidth}
\renewcommand{\arraystretch}{1.58}
  \centering
  \begin{tabular}{|c|c|}
    \cline{1-2}
    Noise model & $\widetilde{\delta}_{_{E}}^{\;\lambda,\,\alpha}$ \\ \hline
    $N_1$ &$+7.0\%$ \\ \hline
    $N_2$ &$+8.8\%$ \\\hline
    $N_3$ &$+11.9\%$ \\ \hline
\end{tabular}
\vspace{0.35cm}
\caption{}
\label{table_hyperparam_tuning}
\end{subfigure}
\vspace{-0.3cm}
\captionsetup{justification=raggedright,singlelinecheck=true}
\caption{Prior hyperparameter tuning results over the phantom set $\mathcal{S}_1$. Fig. \ref{fig_reg_param_tuning}: for noise model $N_1$, regularization parameter $\lambda_{best}$ minimizing $E_{_{MAP}}(\lambda,\alpha_{true})$ in Eq.\eqref{rmse_def_appendix}; similar results are found for $N_2, N_3$. Table \ref{table_hyperparam_tuning}: average relative difference between $E_{_{MAP}}(0.1,\alpha_{CV})$ and $E_{_{MAP}}(\lambda_{best},\alpha_{true})$.}
\label{fig_reg_param_anis_param_tuning}
\end{figure}
In this work, we proceeded as follows.
For each phantom $\mathbf{x}$ in $\mathcal{S}_1$, we initially fix $\alpha=\alpha_{true}$, with $\alpha_{true}$ the anisotropic parameter used to generate that phantom (see Section \ref{dataset}); then, we perform a grid search on $\lambda$, finding the regularization parameter $\lambda_{best}$ that minimizes the RMSE
\begin{equation}\label{rmse_def_appendix}
E_{_{MAP}}(\lambda,\,\alpha) = \sqrt{\frac{1}{N} \sum_{i=0}^{N-1} \bigg( \big(\mathbf{x}\big)_i - \big(\mathbf{x}_{_{MAP}}(\lambda,\, \alpha)\big)_i  \bigg)^2 }
\end{equation}
for $\alpha=\alpha_{true}$, i.e., the RMSE between the ground truth $\mathbf{x}$ and the MAP $\mathbf{x}_{_{MAP}}(\lambda,\, \alpha\!=\!\alpha_{true})$. In Fig.\ref{fig_reg_param_tuning}, we show the distribution of the optimal regularization parameter over $\mathcal{S}_1$, for noise model $N_1$. We obtain similar distributions for noise models $N_2$, $N_3$. Since we find that $E_{_{MAP}}$ is not very sensitive to $\lambda$, we fix $\lambda$ to the average value $\widetilde{\lambda}_{best}=10^{-1}$.
Employing the same regularization parameter for all phantoms allows a fairer comparison of the Bayesian estimates, as they are now obtained from models featuring the same amount of regularization. Moreover, while we selected $\lambda$ by leveraging knowledge of the ground truth for convenience, as it was available for such a phantom study, similar estimates are obtained with approaches \cite{fernandez_vidal_maximum_2020} based on marginal likelihood maximization that rely only upon the noisy tomographic measurements, thus not making use of the ground truth.
For the anisotropic parameter $\alpha$, we use a different approach that we describe next.
Having fixed $\lambda=10^{-1}$, we now assume that $\alpha$ is unknown, and estimate it for each phantom using a cross-validation procedure, here relying only upon the noisy tomographic data $\mathbf{y}\in\mathbb{R}^{100}$. To achieve this, we split the 100 measurements from the lines of sight into 5 randomly chosen sets $\mathbf{y}_{f_i}\in\mathbb{R}^{20},\,i=1,\ldots,5$, with each fold $f_i$ corresponding to 20 lines of sight. Then, we perform a grid search on $\alpha$, with $10^{-4}\leq\alpha\leq10^0$. For each value of $\alpha$ on the chosen grid, we compute five different MAP reconstructions $\mathbf{x}_{_{MAP}}^{f_i},\,i=1,\ldots,5$; each MAP is obtained by employing only 80 measurements, thus hiding the 20 measurements $\mathbf{y}_{f_i}$ corresponding to the $i$-th fold $f_i$. For each MAP $\mathbf{x}_{_{MAP}}^{f_i}$, by applying the forward model $\mathbf{T}$ (see Eq.\eqref{ip_data}) and extracting the entries of interest, we compute the predicted measurements $(\mathbf{T}\,\mathbf{x}_{_{MAP}}^{f_i})_{f_i}$ for the 20 chords not used for reconstruction. By comparing these predictions with the actual measurements $\mathbf{y}_{f_i}$, we compute the mean squared error $\mathrm{MSE}_{\,\alpha,\,f_i}$, measuring the difference between the two. By averaging over the 5 folds, we obtain the metric $\mathrm{MSE}_{\alpha}=1/5(\sum_{i=1}^5\mathrm{MSE}_{\,\alpha,\,f_i})$ used to assess the performance of each $\alpha$. Finally, for the considered phantom, we set the anisotropic parameter to the value $\alpha_{CV}$ that minimizes $\mathrm{MSE}_{\alpha}$. In the described procedure, averaging over 5 different folds increases the robustness of the estimate. Moreover, for each MAP, it is crucial to evaluate the metric on the subset of tomographic data not used for reconstruction. Indeed, it can be shown that the MSE on the chord measurements used for reconstruction decreases monotonically as $\alpha$ decreases: the smaller $\alpha$, the closer the predicted and observed measurements will be, which does not imply, of course, a better reconstruction. The regularization parameter exhibits the same monotonic behavior. To conclude, for the phantom and prior classes considered in this work, selecting $\lambda=10^{-1}$, $\alpha=\alpha_{CV}$ is found to be a sufficient hyperparameter tuning strategy. In Table \ref{table_hyperparam_tuning}, we show the relative difference $\delta_{_E}^{\lambda,\,\alpha}$ between $E_{_{MAP}}(\lambda=10^{-1},\,\alpha_{CV})$ and $E_{_{MAP}}(\lambda_{best},\,\alpha_{true})$, averaged over $\mathcal{S}_1$, showing that the implemented tuning strategy performs acceptably well when compared to the ideal ``tuning'' strategy, consisting in setting $(\lambda, \alpha)=(\lambda_{best},\,\alpha_{true})$ for each phantom.
As a final remark, we point out that, while the proposed strategy performs well within the context of our work, hyperparameter tuning is a complex task, which would benefit from further efforts towards the development of more robust, efficient, and general techniques.\\
Let us now turn our attention to the MCMC hyperparameters $K_b, K$. It is indispensable to choose these well to obtain accurate MC estimates by Eq.\eqref{MC_estimate}. If the number of burn-in iterations $K_b$ is not sufficiently large, the Bayesian estimates are polluted by samples that poorly represent the posterior distribution; if the total number of MCMC iterations $K$ is too small, the estimates will be inaccurate having insufficiently explored the sample space (the MC estimate in Eq.\eqref{MC_estimate} has not converged). We report only the results for noise model $N_1$; noise models $N_2, N_3$ behave similarly. The number of burn-in iterations can be tuned monitoring the log-posterior value: indeed, as mentioned in \cite{fernandez_vidal_maximum_2020}, the log-posterior values of the MCMC samples are expected to stabilize around a plateau value $v_{plat}$ after a few iterations. Such a value is chain-dependent. In Fig.\ref{fig_ula_tuning_logp}, we show the behavior of the normalized log-posterior $\log\,p(\mathbf{x}\vert\mathbf{y}) / v_{plat}\,$, averaged over $\mathcal{S}_1$.
Since by $k=10^3$ iterations the chains have already stabilized around $v_{plat}$, $K_b=10^3$ is a sufficient number of burn-in iterations. Having discarded the first $K_b$ iterations, we must now choose how many iterations $K$ should be performed for MC estimations. Such a choice is subject to a trade-off, since both estimation accuracy and computational cost increase with $K$. To tune $K$, let us run a large number, $10^7$, of MCMC iterations, and compute the resulting mean $\boldsymbol{\mu}_{_{ULA}}^{10^7}$ and standard deviation $\boldsymbol{\sigma}_{_{ULA}}^{10^7}$ using Eq.\eqref{MC_estimate}. Since acquiring even more samples modifies them only marginally, we use $\boldsymbol{\mu}_{_{ULA}}^{10^7}, \boldsymbol{\sigma}_{_{ULA}}^{10^7}$ as ground truth, using them as a benchmark to estimate the error committed by collecting less samples. In particular, let us study the mean absolute percentage errors (MAPEs) $\sum_{i=0}^{N-1}\big\vert (\boldsymbol{\mu}_{_{ULA}}^{10^7})_i - (\boldsymbol{\mu}_{_{ULA}}^{k-K_b})_i \big\vert / (\boldsymbol{\mu}_{_{ULA}}^{10^7})_i\;$, $\sum_{i=0}^{N-1}\big\vert (\boldsymbol{\sigma}_{_{ULA}}^{10^7})_i - (\boldsymbol{\sigma}_{_{ULA}}^{k-K_b})_i \big\vert / (\boldsymbol{\sigma}_{_{ULA}}^{10^7})_i\;$, as a function of $k$. In Figs.\ref{fig_ula_tuning_mape_mean}, \ref{fig_ula_tuning_mape_std}, we plot the MAPEs behavior, averaged over $\mathcal{S}_1$. Since after $k-K_b=10^5$ iterations the MAPEs are typically below the acceptably low value of $2\%$, we decide to use $K=10^5$ iterations for MC estimation.
To conclude, we remark that the selected values $K_b=10^3$, $K=10^5$ are suitable for the posterior distribution in Eq.\eqref{eq_posterior_recon}: such values are not to be expected appropriate for any posterior. The properties of the posterior, which affect those of the Langevin diffusion in Eq.\eqref{langevin_sde}, have a strong influence on the MCMC sampling efficiency. In general, an especially important and desirable property is \emph{geometric ergodicity} \cite{roberts_exponential_1996} of the diffusion process in Eq.\eqref{langevin_sde}. Such a property, which ensures that efficient MCMC sampling based on Eq.\eqref{ULA} can be achieved, is related to the mixing time of the chain: a geometrically ergodic process converges exponentially fast (in time) to the invariant distribution. Convexity of the negative log-posterior, together with the Lipschitz-continuity of its gradient (or of the gradient of its smooth approximation, for the models discussed in Appendix \ref{appendix_prox}), is a sufficient condition guaranteeing geometric ergodicity \cite{pavliotis_stochastic_2014, durmus_nonasymptotic_2017, durmus_efficient_2018}. Therefore, convex imaging models should be preferred, when interested in performing MCMC sampling. The reconstruction model in Eq.\eqref{eq_posterior_recon} is convex, like many other interesting and popular imaging models.

\end{document}